\documentclass[graybox]{svmult}

\usepackage{mathptmx}       % selects Times Roman as basic font
\usepackage{helvet}         % selects Helvetica as sans-serif font
\usepackage{courier}        % selects Courier as typewriter font
\usepackage{type1cm}        % activate if the above 3 fonts are
\usepackage{makeidx}         % allows index generation
\usepackage{graphicx}        % standard LaTeX graphics tool
\usepackage{multicol}        % used for the two-column index
\usepackage[bottom]{footmisc}% places footnotes at page bottom

\usepackage[]{amssymb, amsmath}
\usepackage[]{bm}
\usepackage[]{upgreek}
\usepackage[]{graphicx}
\usepackage[]{nicefrac}
\usepackage[]{subfigure}

\usepackage{algorithm}
\usepackage{algorithmic}

\usepackage[]{xcolor}

\DeclareMathAlphabet{\mathcal}{OMS}{cmsy}{m}{n}
\DeclareMathAlphabet\mathbfcal{OMS}{cmsy}{b}{n}

\DeclareMathOperator*{\maximize}{maximize~}
\DeclareMathOperator*{\minimize}{minimize~}
\DeclareMathOperator{\subjecto}{subject~to~}
\graphicspath{{imgs/}}

\makeindex             % used for the subject index
\begin{document}

\title*{Time-stepping and Krylov methods for large-scale instability problems}
% Use \titlerunning{Short Title} for an abbreviated version of
% your contribution title if the original one is too long
\author{J.-Ch. Loiseau \and M. A. Bucci \and S. Cherubini \and J.-Ch. Robinet}
% Use \authorrunning{Short Title} for an abbreviated version of
% your contribution title if the original one is too long
\institute{J.-Ch. Loiseau \at Laboratoire DynFluid, Arts et M\'etiers ParisTech, 151 boulevard de l'h\^opital, 75013 Paris, France. \email{jean-christophe.loiseau@ensam.eu}
\and M. A. Bucci \at  Laboratoire DynFluid, Arts et M\'etiers ParisTech, 151 boulevard de l'h\^opital, 75013 Paris, France. \email{michele.bucci@ensam.eu}
\and S. Cherubini \at DMMM, Politecnico di Bari, via Re David 200, 70100 Bari, Italy. \email{s.cherubini@gmail.com}
\and J.-Ch. Robinet \at  Laboratoire DynFluid, Arts et M\'etiers ParisTech, 151 boulevard de l'h\^opital, 75013 Paris, France. \email{jean-christophe.robinet@ensam.eu}}
%
% Use the package "url.sty" to avoid
% problems with special characters
% used in your e-mail or web address
%
\maketitle

%%%%%%%%%%%%%%%%%%%%%%%%%%%%
%%%%%                  %%%%%
%%%%%     ABSTRACT     %%%%%
%%%%%                  %%%%%
%%%%%%%%%%%%%%%%%%%%%%%%%%%%

% \abstract*{With the ever increasing computational power available (roughly 20 to 25\% increase annually) and the development of high-performances computing (HPC), investigating the properties of realistic very large-scale nonlinear dynamical systems has been become reachable.}

\abstract{With the ever increasing computational power available and the development of high-performances computing, investigating the properties of realistic very large-scale nonlinear dynamical systems has been become reachable. It must be noted however that the memory capabilities of computers increase at a slower rate than their computational capabilities. Consequently, the traditional matrix-forming approaches wherein the Jacobian matrix of the system considered is explicitly assembled become rapidly intractable. Over the past two decades, so-called \emph{matrix-free} approaches have emerged as an efficient alternative. The aim of this chapter is thus to provide an overview of well-grounded matrix-free methods for fixed points computations and linear stability analyses of very large-scale nonlinear dynamical systems.}

%%%%%%%%%%%%%%%%%%%%%%%%%%%%%%%%%%%%%
%%%%%                           %%%%%
%%%%%     COEUR DU CHAPITRE     %%%%%
%%%%%                           %%%%%
%%%%%%%%%%%%%%%%%%%%%%%%%%%%%%%%%%%%%

% --> Introduction.
\section{Introduction}
\label{sec: introduction}

Simulation of very large-scale linear or non-linear systems is a critical issue in many scientific fields. Fluid dynamics is full of examples where accurate and efficient methods having a reasonable computational cost and memory footprint are required. The study of flow stability is no exception, especially when one is interested in flows where the degree of spatial inhomogeneity is more and more important (one, two or three inhomogeneous directions). Historically, hydrodynamic stability analysis has always evolved according to the progress of computers, but also with the development of increasingly efficient numerical methods.

Before the 1980s, only problems where the flow has a single inhomogeneous spatial direction (generally perpendicular to the advection direction) could be discussed. The first discretization methods used were naturally the spectral or spectral collocations methods \cite{PT83, P2002} which offer a reasonable trade-off between computational cost and resolvability. One of the earliest examples of using such a method for linear stability analysis purposes can be found in \cite{J70}. Approximately at the same time, computation of the eigenvalue spectrum for a 1D flow were carried out \cite{J71, GJ75, M1976}, most often by shooting methods or Newton method coupled with continuation methods \cite{M69, DD1969}. One needs to wait until the mid-70s before eigenvalue solvers based on QR or QZ decompositions \cite{BM1984, BM1987, M90} start to be used in the study of a broad class of flows \cite{O1970, GH1970, MZH85, M90}. With the increased computational power, the 1980s and especially the 90s are marked by the rapid development of these methods for flows of increasing complexity. Various libraries are developed, the most famous ones being LAPACK \cite{LAPACK99}, MKL \cite{intelMKL} and ARPACK \cite{LSY97}. These libraries incorporate many iterative algorithms allowing for the full or partial computation of the eigenspectrum for flows with two inhomogeneous spatial directions, see \cite{T2003,T2011} for a review.

Most of the work carried out during this period consisted of linearizing the governing equations, discretizing them using methods such as spectral methods, finite-differences or finite elements and eventually solving the resulting eigenproblem often with an Arnoldi algorithm \cite{A1951,NO1993,LSY97}. The constant increase in geometric complexity of the flows addressed eventually led to a reformulation of the stability analyses and to the integration of these methods into existing simulation codes (e.g.\ FreeFem++ \cite{H2012}, Nek5000 \cite{FKMTLK08} or Nektar++ \cite{KS2005}). This evolution led to the increase importance of the numerical part (which was initially of theoretical nature). A glaring example of the weight of the numerics and resolution methods for very large-scale nonlinear dynamical systems can be illustrated in the computation of base flows, fixed point of the governing equations, which, unlike parallel and geometrically simple flows, can no longer be analytically obtained or simply approximated. Accurate computation of these equilibrium solutions is thus necessary. Fixed points solvers such as the selective frequency damping \cite{pof:akervik:2006}, Newton \cite{PW1998} and quasi-Newton \cite{SB2002}, or more recently RPM (Recursive Projection Method) \cite{siam:shroff:1993} and Boostconv \cite{citro2017efficient} are now commonly used to compute these equilibrium solutions.

Regarding the computation of the eigenpairs of the linearized Navier-Stokes operator, different strategies have been proposed over the years. When one tries to compute the stability of a fully three-dimensional flow, the computation and the manipulation of the Jacobian operator is a key problem mainly related to its dimension, of the order $10^6$-$10^8$. In the literature, two major approaches have emerged. The first one, known as "matrix-forming", explicitly assembles the Jacobian matrix The advantage of such an approach is that it is simple to compute the adjoint operator, which in this case is the hermitian of the discrete Jacobian matrix. However, this approach currently runs into computational difficulties for three-dimensional flows. Indeed, eigenvalue solvers typically require the computation of the inverse of the Jacobian, whose computational cost becomes almost out of reach. In the second approach, called "matrix-free", the Jacobian matrix is not explicitly assembled. Instead, one only needs to be able to evaluate the matrix-vector product so as to generate a Krylov sequence from which the spectral properties of the Jacobian are approximated. This method has the advantage of making stability analyses of very large-scale problems doable. One of its major drawbacks however is that one needs to write the continuous adjoint equations if interested into receptivity, sensitivity or non-modal stability problems.

The aim of this chapter is to take the point of view of the latter approach and to describe the main principles for both modal and non-modal analyses within a matrix-free and time-stepper computational framework. In that aspect, it follows the works of \cite{tuckerman2000bifurcation} and \cite{dijkstra2014numerical}. The different algorithms enabling the computation of the fixed points and the analysis of their modal and non-modal stability properties will be presented in detail. Advantages and limitations of each method will also be presented and illustrated by simple examples. The second objective is to give the reader a guide on how to use the different methods in order to implement them into an existing CFD code. For that purpose, the chapter is organized as follows: first, the theoretical frameworks of fixed points computation and modal and non-modal stability analyses are presented. The other sections present the different algorithms one needs to use for such analyses, taking care to compare their performances and to illustrate them on representative cases. Finally, the chapter ends with a conclusion and perspectives highlighting the most recent evolution of these methods and their possible extensions to more complex dynamics, especially to very large-scale time-periodic nonlinear dynamical systems.

% --> Theoretical framework.
\section{Theoretical framework}
\label{sec: theory}

  Our attention is focused on the characterization of very high-dimensional nonlinear dynamical systems typically arising from the spatial discretization of partial differential equations such as the incompressible Navier-Stokes equations. In general, the resulting dynamical equations are written down as a system of firt order differential equations
  \begin{equation}
    \dot{X}_j = \mathcal{F}_j \left( \left\{ X_i(t);\ i =1, \cdots, n \right\}, t \right)
    \notag
  \end{equation}
  where the integer $n$ is the \emph{dimension} of the system, and $\dot{X}_j$ denotes the time-derivative of $X_j$. Using the notation $\bf{X}$ and $\mathbfcal{F}$ for the sets $\left\{ X_j,\ i =1, \cdots, n \right\}$ and $\left\{ \mathcal{F}_j,\ i =1, \cdots, n \right\}$, this system can be compactly written as
  \begin{equation}
    \dot{\mathbf{X}} = \mathbfcal{F}(\mathbf{X}, t),
    \label{eq: theory -- continuous-time dynamical system}
  \end{equation}
  where $\mathbf{X}$ is the $n \times 1$ \emph{state vector} of the system and $t$ is a continuous variable denoting time. Alternatively, accounting also for temporal discretization gives rise to a discrete-time dynamical system
  \begin{equation}
    X_{j, k+1} = \mathcal{G}_j \left( \left\{ X_{i, k};\ i = 1, \cdots, n \right\}, k \right)
    \notag
  \end{equation}
  or formally
  \begin{equation}
    \mathbf{X}_{k+1} = \mathbfcal{G} \left( \mathbf{X}_k, k \right)
    \label{eq: theory -- discrete-time dynamical system}
  \end{equation}
  where the index $k$ now denotes the discrete time variable. If one uses first-order Euler extrapolation for the time discretization, the relation between $\mathbfcal{F}$ and $\mathbfcal{G}$ is given by
  \begin{equation}
    \mathbfcal{G} (\mathbf{X}) = \mathbf{X} + \Updelta t \mathbfcal{F}\left( \mathbf{X} \right),
    \notag
  \end{equation}
  where $\Updelta t$ is the time-step and the explicit dependences on $t$ and $k$ have been dropped for the sake of simplicity.

  In the rest of this section, the reader will be introduced to the concepts of fixed points and linear stability, two concepts required to characterize a number of properties of the system investigated. Particular attention will be paid to \emph{modal} and \emph{non-modal stability}, two approaches that have become increasingly popular in fluid dynamics over the past decades. Note that the concept of \emph{nonlinear optimal perturbation}, which has raised a lot attention lately, is beyond the scope of the present contribution. For interested readers, please refer to the recent work by \cite{nonlinear_optimal:kerswell:2014} and references therein.

  Finally, while we will mostly use the continuous-time representation \eqref{eq: theory -- continuous-time dynamical system} when introducing the reader to the theoretical concepts exposed in this section, using the discrete-time representation \eqref{eq: theory -- discrete-time dynamical system} will prove more useful when discussing and implementing the different algorithms presented in \textsection \ref{sec: numerics}.

  %%%%%%%%%%%%%%%%%%%%%%%%%%%%%%%%
  %%%%%                      %%%%%
  %%%%%     FIXED POINTS     %%%%%
  %%%%%                      %%%%%
  %%%%%%%%%%%%%%%%%%%%%%%%%%%%%%%%

  \subsection{Fixed points}
  \label{subsec: theory-fixed points}

  Nonlinear dynamical systems described by Eq.~\eqref{eq: theory -- continuous-time dynamical system} or Eq.~\eqref{eq: theory -- discrete-time dynamical system} tend to admit a number of different equilibria forming the backbone of their phase space. These different equilibria can take the form of fixed points, periodic orbits, torus or strange attractors for instance. In the rest of this work, our attention will be solely focused on fixed points.

\begin{figure}[b]
    \centering
    \sidecaption
    \includegraphics[scale=1]{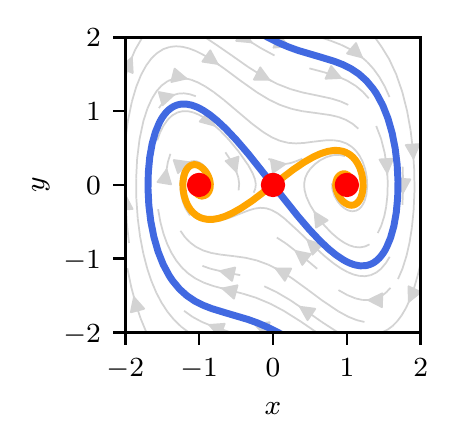}
    \caption{Phase portrait of the unforced Duffing oscillator \eqref{eq: theory -- Duffing oscillator}. The red dots denote the three fixed points admitted by the system. The blue (resp. orange) thick line depicts the stable (resp. unstable) manifold of the saddle point located at the origin. Grey lines highlight a few trajectories exhibited for different initial conditions.}
    \label{fig: theory -- Duffing oscillator}
  \end{figure}

  For a continuous-time dynamical system described by Eq.~\eqref{eq: theory -- continuous-time dynamical system}, fixed points $\mathbf{X}^{*}$ are solution to
  \begin{equation}
    \mathbfcal{F}\left( \mathbf{X} \right) = 0.
    \label{eq: theory -- continuous-time fixed point}
  \end{equation}
  Conversely, fixed points $\mathbf{X}^*$ of a discrete-time dynamical system described by Eq.~\eqref{eq: theory -- discrete-time dynamical system} are solution to
  \begin{equation}
    \mathbfcal{G} \left( \mathbf{X} \right) = \mathbf{X}.
    \label{eq: theory -- discrete-time fixed point}
  \end{equation}
  It must be emphasized that both Eq.~\eqref{eq: theory -- continuous-time fixed point} and Eq.~\eqref{eq: theory -- discrete-time fixed point} may admit multiple solutions. Such a multiplicity of fixed points can easily be illustrated by a dynamical system as simple as the following Duffing oscillator
  \begin{equation}
    \begin{aligned}
      \dot{x} & = y \\
      \dot{y} & = -\displaystyle \frac{1}{2} y + x - x^3.
    \end{aligned}
    \label{eq: theory -- Duffing oscillator}
  \end{equation}
  Despite its apparent simplicity, this Duffing oscillator admits three fixed points, namely
  \begin{itemize}
    \item a saddle at the origin $\mathbf{X}^* = (0, 0)$,
    \item two linearly stable spirals located at $\mathbf{X}^* = (\pm 1, 0)$.
  \end{itemize}
  All of these fixed points, along with some trajectories, are depicted on figure \ref{fig: theory -- Duffing oscillator} for the sake of illustration. Such a multiplicity of fixed points also occurs in dynamical systems as complex as the Navier-Stokes equations. Determining which of these fixed points is the most relevant one from a physical point of view is problem-dependent and left for the user to decide. Note however that computing these equilibrium points is a prerequisite to all of the analyses to be described in this chapter. Numerical methods to solve Eq.~\eqref{eq: theory -- continuous-time fixed point} or Eq.~\eqref{eq: theory -- discrete-time fixed point} will be discussed in \textsection \ref{subsec: numerics-fixed points computation}.

  %%%%%%%%%%%%%%%%%%%%%%%%%%%%%%%%%%%%
  %%%%%                          %%%%%
  %%%%%     LINEAR STABILITY     %%%%%
  %%%%%                          %%%%%
  %%%%%%%%%%%%%%%%%%%%%%%%%%%%%%%%%%%%

  \subsection{Linear stability analysis}
  \label{subsec: theory -- linear stability}

  Having computed a given fixed point $\mathbf{X}^*$ of a continuous-time nonlinear dynamical system given by Eq. \eqref{eq: theory -- continuous-time dynamical system}, one may ask whether it corresponds to a stable or unstable equilibrium of the system. Before pursuing, the very notion of \emph{stability} needs to be explained. It is traditionally defined following the concept of Lyapunov stability. Having computed the equilibrium state $\mathbf{X}^*$, the system is perturbed around this state. If it returns back to the equilibrium point, the latter is deemed stable, otherwise, it is regarded as unstable. It has to be noted that, in the concept of Lyapunov stability, an infinite time horizon is allowed for the return to equilibrium.

  The dynamics of a perturbation $\mathbf{x} = \mathbf{X} - \mathbf{X}^*$ are governed by
  \begin{equation}
    \dot{\mathbf{x}} = \mathbfcal{F}(\mathbf{X}^* + \mathbf{x}).
  \end{equation}
  Assuming the perturbation $\mathbf{x}$ is infinitesimal, $\mathbfcal{F}(\mathbf{X})$ can be approximated by its first-order Taylor expansion around $\mathbf{X} = \mathbf{X}^*$. Doing so, the governing equations for the perturbation $\mathbf{x}$ simplify to
  \begin{equation}
    \dot{\mathbf{x}} = \mathbfcal{A}\mathbf{x},
    \label{eq: theory -- linear perturbation dynamics}
  \end{equation}
  where $\mathbfcal{A}=\partial\mathbfcal{F}/\partial\mathbf{X}$ is the $n \times n$ Jacobian matrix of $\mathbfcal{F}$. Starting from an initial condition $\mathbf{x}_0$, the perturbation at time $t$ is given by
  \begin{equation}
    \mathbf{x}(t) = \exp \left( \mathbfcal{A}t \right) \mathbf{x}_0.
    \label{eq: theory -- linear stability solution}
  \end{equation}
  The operator $\mathbfcal{M}(t) = \exp \left( \mathbfcal{A}t \right)$ is known as the \emph{exponential propagator}. Introducing the spectral decomposition of $\mathbfcal{A}$
  \begin{equation}
    \mathbfcal{A} = \mathbfcal{V} \boldsymbol{\Lambda} \mathbfcal{V}^{-1},
    \notag
  \end{equation}
  Eq. \eqref{eq: theory -- linear stability solution} can be rewritten as
  \begin{equation}
    \mathbf{x}(t) = \mathbfcal{V} \exp \left( \boldsymbol{\Lambda} t \right) \mathbfcal{V}^{-1} \mathbf{x}_0,
  \end{equation}
  where the i\textsuperscript{th} column of $\mathbfcal{V}$ is the eigenvector $\mathbf{v}_i$ associated to the i\textsuperscript{th} eigenvalue $\lambda_i = \boldsymbol{\Lambda}_{ii}$, with $\boldsymbol{\Lambda}$ a diagonal matrix. Assuming that the eigenvalues of $\mathbfcal{A}$ have been sorted by decreasing real part, it can easily be shown that
  \begin{equation}
    \lim\limits_{t \to + \infty} \exp \left( \mathbfcal{A} t \right) \mathbf{x}_0 = \lim \limits_{t \to + \infty} \exp \left( \lambda_1 t \right) \mathbf{v}_1 .
    \notag
  \end{equation}
  The asymptotic fate of an initial perturbation $\mathbf{x}_0$ is thus entirely dictated by the real part of the leading eigenvalue $\lambda_1$:
  \begin{itemize}
    \item if $\Re \left( \lambda_1 \right) > 0$, a random initial perturbation $\mathbf{x}_0$ will eventually grow exponentially rapidly. Hence, the fixed point $\mathbf{X}^*$ is deemed \emph{linearly unstable}.

    \item If $\Re \left( \lambda_1 \right) < 0$, the initial perturbation $\mathbf{x}_0$ will eventually decay exponentially rapidly. The fixed point $\mathbf{X}^*$ is thus \emph{linearly stable}.
  \end{itemize}
  The case $\Re \left( \lambda_1 \right) = 0$ is peculiar. The fixed point $\mathbf{X}^*$ is called \emph{elliptic} and one cannot conclude about its stability solely by looking at the eigenvalues of $\mathbfcal{A}$. In this case, one needs to resort to \emph{weakly non-linear analysis} which essentially looks at the properties of higher-order Taylor expansion of $\mathbfcal{F} \left( \mathbf{X} \right)$. Once again, this is beyond the scope of the present chapter. Interested readers are referred to \cite{SS1971} for more details about such analyses.

  \paragraph*{Illustration}

\begin{figure}[b]
    \centering
    \sidecaption
    \includegraphics[scale=1]{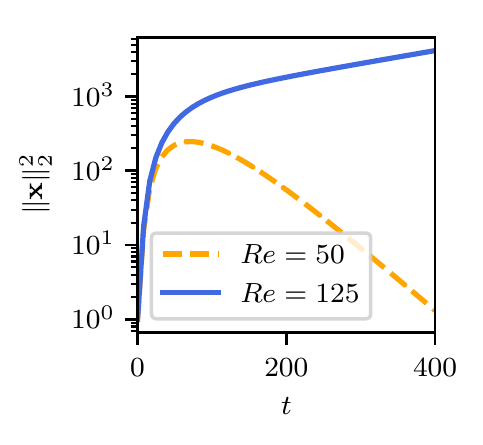}
    \caption{Evolution as a function of time of $\| \mathbf{x} \|_2^2 = x_1^2 + x_2^2$ for the toy-model \eqref{eq: theory -- schmid system}. For $Re=50$ (resp. $Re=125$), the asymptotic fate of $\| \mathbf{x} \|_2^2$ is described by a decreasing (resp. increasing) exponential. For $Re=50$, the equilibrium point is thus linear stable, while it is linearly unstable for $Re=125$.}
    \label{fig: theory -- illustration modal stability}
  \end{figure}

  Let us illustrate the notion of linear stability on a simple example. For that purpose, we will consider the same linear dynamical system as in \cite{amr:schmid:2014}. This system reads
  \begin{equation}
    \displaystyle \frac{\mathrm{d}}{\mathrm{d}t} \begin{bmatrix} x_1 \\ x_2 \end{bmatrix} =
    \underbrace{
    \begin{bmatrix}
      \displaystyle \frac{1}{100} - \frac{1}{Re} & 0 \\
      1 & \displaystyle -\frac{2}{Re}
    \end{bmatrix}
    }_{\mathbfcal{A}}
    \begin{bmatrix} x_1 \\ x_2 \end{bmatrix}
    \label{eq: theory -- schmid system}
  \end{equation}
  where $Re$ is a control parameter. For such a simple case, it is obvious that the eigenvalues of $\mathbfcal{A}$ are given by
  \begin{equation}
    \lambda_1 = \displaystyle \frac{1}{100} - \frac{1}{Re}
    \notag
  \end{equation}
  and
  \begin{equation}
    \lambda_2 = - \frac{2}{Re}.
    \notag
  \end{equation}
  While $\lambda_2$ is constantly negative, $\lambda_1$ is negative for $Re < 100$ and positive otherwise. Figure \ref{fig: theory -- illustration modal stability} depicts the time-evolution of $\| \mathbf{x} \|_2^2 = x_1^2 + x_2^2$ for two different values of $Re$. Please note that the short-time ($t < 100$) behavior of the perturbation will be discussed in \textsection \ref{subsec: theory -- non-modal stability}. It is clear nonetheless that, for $t>100$, the time-evolution of the perturbation can be described by an exponential function. Whether this exponential increases or decreases as a function of time is solely dictated by the sign of $\lambda_1$, negative for $Re=50$ and positive for $Re=100$. For $Re=50$, the equilibrium point $\mathbf{X}^* = \begin{bmatrix} 0 & 0 \end{bmatrix}^T$ is thus stable, while it is unstable for $Re=125$.

  %%%%%%%%%%%%%%%%%%%%%%%%%%%%%%%%%%%%%%%%%%%%%%%%
  %%%%%                                      %%%%%
  %%%%%     NON-MODAL STABILITY ANALYSIS     %%%%%
  %%%%%                                      %%%%%
  %%%%%%%%%%%%%%%%%%%%%%%%%%%%%%%%%%%%%%%%%%%%%%%%

  \subsection{Non-modal stability analysis}
  \label{subsec: theory -- non-modal stability}

\begin{figure}[b]
    \centering
    \includegraphics[scale=1]{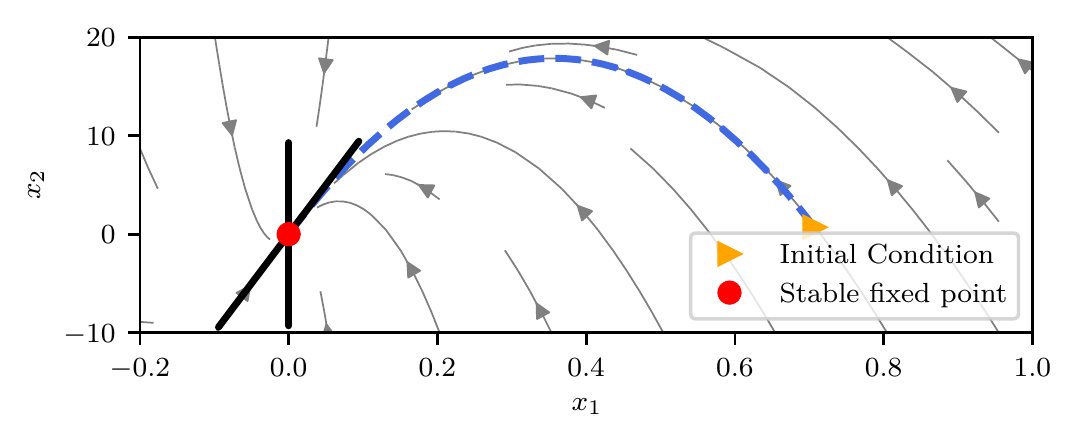}
    \caption{The blue (dashed) line shows the trajectory stemming from a random unit-norm initial condition $\mathbf{x}_0$. The thick black lines depict the two linear manifolds of the fixed point. The diagonal one corresponds to $\lambda_1 = \nicefrac{1}{100} - \nicefrac{1}{Re}$, while the vertical one is associated to $\lambda_2 = -\nicefrac{2}{Re}$. In the present case, $Re$ is set to 50, thus corresponding to a situation where the fixed point is linearly stable.}
    \label{fig: theory -- illustration transient growth}
  \end{figure}

  Looking once more at figure \ref{fig: theory -- illustration modal stability}, it can be seen that, although the system is linearly stable for $Re=50$, the perturbation $\mathbf{x}$ can experience a transient growth of its energy for a short period of time, roughly given by $0 < t <100$ in the present case, before its eventual exponential decay. This behavior is related to the \emph{non-normality} of $\mathbfcal{A}$, i.e.\
  \begin{equation}
    \mathbfcal{A}^{\dagger} \mathbfcal{A} \neq \mathbfcal{A} \mathbfcal{A}^{\dagger},
    \label{eq: theory -- non-normality equation}
  \end{equation}
  where $\mathbfcal{A}^{\dagger}$ is the \emph{adjoint} of $\mathbfcal{A}$. As a result of this non-normality, the eigenvectors of $\mathbfcal{A}$ do not form an orthonormal set of vectors\footnote{Note that the non-normality of $\mathbfcal{A}$ also implies that its right and left eigenvectors are different. This observation may have large consequences in fluid dynamics, particularly when addressing the problems of optimal linear control and/or estimation of strongly non-parallel flows.}. The consequences of this non-orthogonality of the set of eigenvectors can be visualized on figure \ref{fig: theory -- illustration transient growth} where the trajectory stemming from a random unit-norm initial condition $\mathbf{x}_0$ is depicted in the phase plane of our toy-model \eqref{eq: theory -- schmid system}. The perturbation $\mathbf{x}(t)$ is first attracted toward the linear manifold associated to the least stable eigenvalue $\lambda_1$, causing in the process the transient growth of its energy by a factor 300. Once it reaches the vicinity of the linearly stable manifold, the perturbation eventually decays exponentially rapidly along this eigendirection of the fixed point. The next sections are devoted to the introduction of mathematical tools particularly useful to characterize phenomena resulting from this non-normality of $\mathbfcal{A}$, both in the time and frequency domains, when the fixed point considered is stable.

    %-----> Optimal perturbation.
    \subsubsection{Optimal perturbation analysis}
    \label{subsubsec: theory -- optimal perturbation}

    Having observed that a random initial condition can experience a relatively large transient growth of its energy over a short period of time even though the fixed point is stable, one may be interested in the worst case scenario, i.e.\ finding which initial condition $\mathbf{x}_0$ is amplified as much as possible before it eventually decays. Searching for such a perturbation is known as \emph{optimal perturbation analysis} and can be addressed by two different methods:
    \begin{itemize}
      \item Optimization,
      \item Singular Value Decomposition (SVD).
    \end{itemize}
    Both approaches will be presented. Although it requires the introduction of additional mathematical concepts, the approach relying on optimization will be introduced first in \textsection \ref{paragraph: theory -- optimal perturbation optimization} as it is easier to grasp. The approach relying on singular value decomposition of the exponential propagrator $\mathbfcal{M} = \exp \left( \mathbfcal{A} t \right)$ will then be presented in \textsection \ref{paragraph: theory -- optimal perturbation svd}.

      % --> Lagrange multipliers.
      \paragraph{Formulation as an optimization problem}
      \label{paragraph: theory -- optimal perturbation optimization}

      The aim of optimal perturbation analysis is to find the unit-norm initial condition $\mathbf{x}_0$ that maximizes $\| \mathbf{x}(T) \|_2^2$, where $T$ is known as the \emph{target time}. Note that we here consider only the 2-norm of $\mathbf{x}(T)$ for the sake of simplicity, although one could formally optimize different norms, see \cite{FCS2012, jfm:foures:2013, jfm:foures:2014, fdr:farano:2016} for examples from fluid dynamics. For a given target time $T$, such a problem can be formulated as the following constrained maximization problem
      \begin{equation}
          \begin{aligned}
            \maximize \limits_{\mathbf{x}_0} & \mathcal{J} \left( \mathbf{x}_0 \right) = \| \mathbf{x}(T) \|_2^2\\
            \subjecto & \dot{\mathbf{x}} - \mathbfcal{A}\mathbf{x} = 0 \\
            ~ & \| \mathbf{x}_0 \|_2^2 - 1 = 0,
          \end{aligned}
          \label{eq: theory -- constrained maximization}
      \end{equation}
      where $\mathcal{J}(\mathbf{x}_{0})$ is known as the \emph{objective function}. It must be emphasized that problem \eqref{eq: theory -- constrained maximization} is not formulated as a convex optimization problem\footnote{
      Formally, a convex optimization problem reads
      \begin{equation}
        \begin{aligned}
          \minimize \limits_{\mathbf{x}} & \mathcal{J} \left( \mathbf{x} \right) \\
          \subjecto & g_i \left( \mathbf{x} \right) \leq 0, \ i = 1, \cdots, m \\
          ~ & h_i \left( \mathbf{x} \right) = 0, \ i = 1, \cdots, p,
        \end{aligned}
        \notag
      \end{equation}
      where the objective function $\mathcal{J} \left( \mathbf{x} \right)$ and the inequality constraints functions $g_i \left( \mathbf{x} \right)$ are convex. The conditions on the equality constraints functions $h_i \left( \mathbf{x} \right)$ are more restrictive as they need to be affine functions, i.e.\ of the form $h_i \left( \mathbf{x} \right) = \mathbf{a}_i^T \mathbf{x} + b_i$. See the book by Boyd \& Vandenberghe \cite{book:boyd:2004} for extensive details about convex optimization.}.
      As such, it may exhibit local maximum. Nonetheless, this constrained maximization problem can be recast into the following unconstrained maximization problem
      \begin{equation}
        \maximize \limits_{\mathbf{x}, \mathbf{v}, \mu} \mathcal{L} \left( \mathbf{x}, \mathbf{v}, \mu \right),
        \label{eq: theory -- unconstrained maximization}
      \end{equation}
      where
      \begin{equation}
        \mathcal{L} \left( \mathbf{x}, \mathbf{v}, \mu \right) = \mathcal{J}\left( \mathbf{x}_0 \right) + \int_{0}^T \mathbf{v}^T \left( \dot{\mathbf{x}} - \mathbfcal{A}\mathbf{x} \right) \mathrm{d}t + \mu \left( \| \mathbf{x}_0 \|_2^2 - 1 \right)
        \label{eq: theory -- augmented Lagrangian}
      \end{equation}
      is known as the \emph{augmented Lagrangian} function. The additional optimization variables $\mathbf{v}$ and $\mu$ appearing in the definition of the augmented Lagrangian $\mathcal{L}$ are called \emph{Lagrange multipliers}. Solutions to problem \eqref{eq: theory -- unconstrained maximization} are identified by vanishing first variations of $\mathcal{L}$ with respect to our three optimization variables. The first variation of $\mathcal{L}$ with respect to $\mathbf{v}$ and $\mu$ are simply the constraints of our original problem \eqref{eq: theory -- constrained maximization}. The first variation of $\mathcal{L}$ with respect to $\mathbf{x}$ on the other hand is given by
      \begin{equation}
        \delta_{\mathbf{x}} \mathcal{L} = \left[ \nabla_{\mathbf{x}} \mathcal{J} + \mathbf{v}(T) \right] \cdot \delta \mathbf{x}(0) + \int_0^T \left[ \dot{\mathbf{v}} - \mathbfcal{A}^{\dagger} \mathbf{v} \right] \cdot \delta \mathbf{x} \ \mathrm{dt} + \left[ 2\mu \mathbf{x}_0 - \mathbf{v}(0) \right] \cdot \delta \mathbf{x}(0).
        \label{eq: theory -- optimality condition}
      \end{equation}
      Eq. \eqref{eq: theory -- optimality condition} vanishes only if
      \begin{equation}
        \dot{\mathbf{v}} = \mathbfcal{A}^{\dagger} \mathbf{v} \ \text{ over } t \in \left( 0, T \right),
        \label{eq: theory -- adjoint equations}
      \end{equation}
      and
      \begin{equation}
        \begin{aligned}
          \nabla_{\mathbf{x}} \mathcal{J} - \mathbf{v}(T) & = 0 \\
          2\mu \mathbf{x}_0 - \mathbf{v}(0) & = 0.
        \end{aligned}
        \label{eq: theory -- compatibility conditions}
      \end{equation}
      Note that Eq. \eqref{eq: theory -- adjoint equations} is known as the adjoint system\footnote{
      Given an appropriate inner product, the adjoint operator $\mathbfcal{A}^{\dagger}$ is defined such that
      \begin{equation}
        \langle \mathbf{v} \vert \mathbfcal{A} \mathbf{x} \rangle = \langle \mathbfcal{A}^{\dagger} \mathbf{v} \vert \mathbf{x} \rangle,
        \notag
      \end{equation}
      where $\langle \mathbf{a} \vert \mathbf{b} \rangle$ denotes the inner product of $\mathbf{a}$ and $\mathbf{b}$. If one consider the classical Euclidean inner product, the adjoint operator is simply given by
      $$\mathbfcal{A}^{\dagger} = \mathbfcal{A}^H$$
      where $\mathbfcal{A}^H$ is the Hermitian (i.e.\ complex-conjugate transpose) of $\mathbfcal{A}$. It must be noted finally that the direct operator $\mathbfcal{A}$ and the adjoint one $\mathbfcal{A}^{\dagger}$ have the same eigenspectrum. This last observation is a key point when one aims at validating the numerical implementation of an adjoint solver.
      }
      of our original linear dynamical system, while Eq. \eqref{eq: theory -- compatibility conditions} are called compatibility conditions. Maximizing $\mathcal{L}$ is then a problem of simultaneously satisfying \eqref{eq: theory -- linear perturbation dynamics}, \eqref{eq: theory -- adjoint equations} and \eqref{eq: theory -- compatibility conditions}. This is in general done iteratively by gradient-based algorithms such as gradient ascent or the rotation-update gradient algorithm (see \textsection \ref{sec: numerics}). For more details about adjoint-based optimization, see \cite{book:boyd:2004, nonlinear_optimal:kerswell:2014}.

      %--> A Rayleigh quotient problem.
      \paragraph{Formulation using SVD}
      \label{paragraph: theory -- optimal perturbation svd}

      As stated previously, formulating the optimal perturbation analysis as a constrained maximization results in a non-convex optimization problem \eqref{eq: theory -- constrained maximization}. Consequently, although a solution to \eqref{eq: theory -- constrained maximization} can easily be obtained by means of gradient-based algorithms, one cannot rule out the possibility that this solution is only a local maximum rather than the global one. In this section, we will show that recasting problem \eqref{eq: theory -- constrained maximization} in the framework of linear algebra however allows us to obtain easily this global optimal.

      Let us first redefine our optimization problem as
      \begin{equation}
        \maximize_{\mathbf{x}_0} \displaystyle \frac{\| \mathbf{x}(T) \|_2^2}{\| \mathbf{x}_0 \|_2^2}
        \label{eq: theory -- energy gain}
      \end{equation}
      so that rather than maximizing $\| \mathbf{x}(T) \|_2^2$ under the constraint that $\| \mathbf{x}_0 \|_2^2 = 1$, we now directly aim to maximize the energy gain $\mathcal{G}(T) = \nicefrac{\| \mathbf{x}(T) \|_2^2}{\| \mathbf{x}_0 \|_2^2}$. Moreover, recalling from \eqref{eq: theory -- linear stability solution} that
      \begin{equation}
        \mathbf{x}(T) = \exp \left( \mathbfcal{A} T \right) \mathbf{x}_0,
        \notag
      \end{equation}
      our energy gain maximization problem can finally be written as
      \begin{equation}
        \begin{aligned}
          \mathcal{G}(T) & = \max_{\mathbf{x}_0} \displaystyle \frac{\| \exp \left( \mathbfcal{A} t \right) \mathbf{x}_0 \|_2^2}{\| \mathbf{x}_0 \|_2^2} \\
          & = \| \exp \left( \mathbfcal{A} T \right) \|_2^2
        \end{aligned}
      \end{equation}
      where $\| \exp \left( \mathbfcal{A} T \right) \|_2$ is a vector-induced matrix norm taking care of the optimization over all possible initial conditions $\mathbf{x}_0$. Introducing singular value decomposition (SVD), i.e.\
      \begin{equation}
        \mathbfcal{M} = \mathbfcal{U} \boldsymbol{\Sigma} \mathbfcal{V}^H,
        \notag
      \end{equation}
      it is relatively easy to demonstrate that the optimal energy gain $\mathcal{G}(T)$ is given by
      \begin{equation}
        \mathcal{G}(T) = \sigma_1^2,
        \label{theory -- optimal energy gain }
      \end{equation}
      where $\sigma_1$ is the largest singular value of the exponential propagator $\mathbfcal{M} = \exp \left( \mathbfcal{A} T \right)$. The optimal initial condition $\mathbf{x}_0$ is then given by the principal right singular vector (i.e.\ $\mathbf{x}_0 = \mathbf{v}_1$), while the associated response is given by $\mathbf{x}(T) = \sigma_1 \mathbf{u}_1$, where $\mathbf{u}_1$ is the principal left singular vector.

      % --> Illustration.
      \paragraph{Illustration}

\begin{figure}[b]
        \centering
        \includegraphics[width=\textwidth]{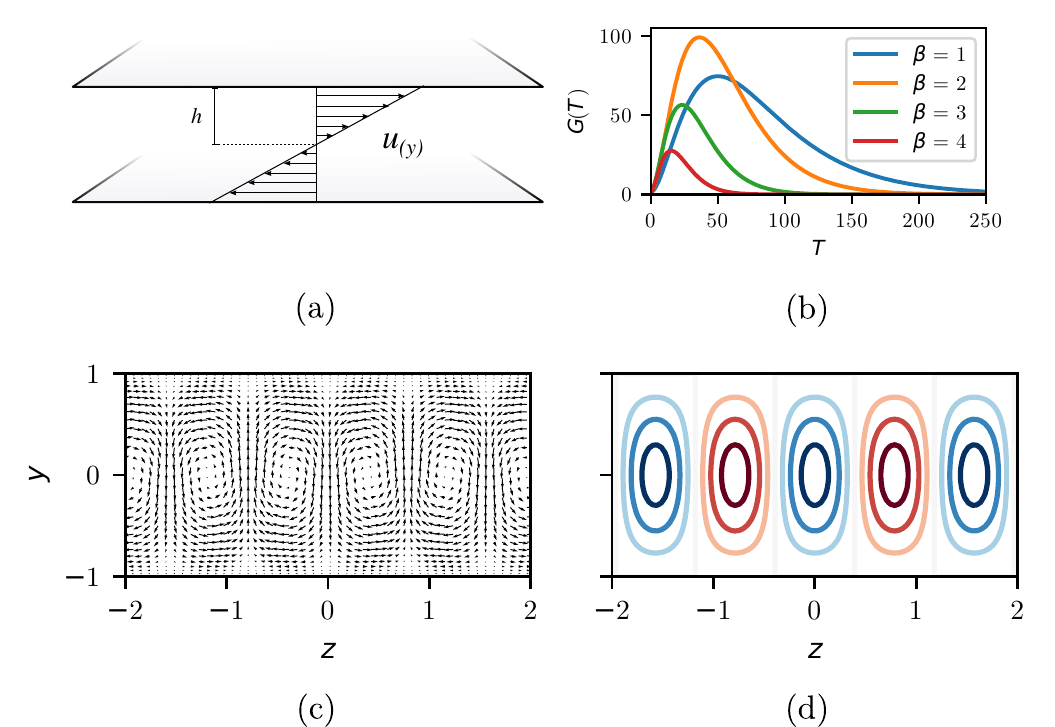}
        \caption{Illustration of optimal perturbation analysis for the plane Couette flow at $Re=300$. In all cases, the streamwise wavenumber of the perturbation is set to $\alpha=0$. (a) Optimal gain curve for different spanwise wavenumbers $\beta$. (b) Optimal perturbation (left) and optimal response (right) for $\beta=2$. Note that optimal perturbation consists of streamwise oriented vortices, while the associated response at time $T$ consist in high- and low-speed streaks.}
        \label{fig: theory -- optimal perturbation illustration}
      \end{figure}

      As to illustrate linear optimal perturbations, let us consider the incompressible flow of a Newtonian fluid induced by two flat plates moving in-plane in opposite directions as sketched on figure \ref{fig: theory -- optimal perturbation illustration}(a). The resulting flow, known as \emph{plane Couette flow}, is given by
      $$U(y) = y.$$
      Note that it is a linearly stable fixed point of the Navier-Stokes equations no matter the Reynolds number considered. Despite its linear stability, subcritical transition to turbulence can occur for Reynolds numbers as low as $Re=325$ \cite{M2016}.

      Without getting too deep into the mathematical and physical details of such subcritical transition, part of the explanation can be given by linear optimal perturbation analysis. The dynamics of an infinitesimal perturbation $\mathbf{x} = \begin{bmatrix} \mathbf{v} & \mathbf{\eta} \end{bmatrix}^T$, characterized by a certain wavenumber $\mathbf{k} = \alpha \mathbf{e}_x + \beta \mathbf{e}_z$, evolving in the vicinity of this fixed point are governed by
      \begin{equation}
        \displaystyle \frac{\mathrm{d}}{\mathrm{d}t} \begin{bmatrix} \mathbf{v} \\ \mathbf{\eta} \end{bmatrix} = \begin{bmatrix} \mathbfcal{A}_{OS} & 0 \\ \mathbfcal{C} & \mathbfcal{A}_{S} \end{bmatrix} \begin{bmatrix} \mathbf{v} \\ \mathbf{\eta} \end{bmatrix}
        \label{eq: theory -- orr-sommerfeld-squire equations}
      \end{equation}
      where $\mathbf{v}$ is the wall-normal velocity of the perturbation and $\mathbf{\eta}$ its wall-normal vorticity, $\mathbfcal{A}_{OS}$ is the Orr-Sommerfeld operator, while $\mathbfcal{A}_{S}$ is the Squire one. The operator $\mathbfcal{C}$ describes the existing coupling between the wall-normal velocity $\mathbf{v}$ and the wall-normal vorticity $\mathbf{\eta}$. For certain pairs of wavenumbers, this Orr-Sommerfeld-Squire operator is highly non-normal and perturbations can exhibit very large transient growth. This is illustrated on figure \ref{fig: theory -- optimal perturbation illustration}(a) where the evolution of the optimal gain $\mathcal{G}(T)$ as a function of the target time $T$ is depicted for different pairs of wavenumbers $(\alpha, \beta)$ at $Re=300$. The maximum amplification achievable over all target times $T$ and wavenumbers pairs $(\alpha, \beta)$ is $\mathcal{G}_{\mathrm{opt}} \simeq 100$. The initial perturbation $\mathbf{x}_0$ corresponding to this optimal energy gain is depicted on figure \ref{fig: theory -- optimal perturbation illustration}(b). It corresponds to streamwise-oriented vortices that eventually give rise to streamwise velocity streaks due to the lift-up effect \cite{jfm:landahl:1980, ejmbf:brandt:2014}, see figure \ref{fig: theory -- optimal perturbation illustration}(b). While this perturbation eventually decays exponentially rapidly in a purely linear framework, it has been shown that, even for a moderately large initial amplitude, it may eventually trigger transition to turbulence when used as initial condition in a non-linear direct numerical simulation of the Navier-Stokes equations \cite{LB2014}. For more details about subcritical transition and extension of optimal perturbation analysis to non-linear operators, interested readers are referred to \cite{K2018}.

    \subsubsection{Resolvent analysis}
    \label{subsubsec: theory -- resolvent perturbation}

    The optimal perturbation analysis (see \textsection \ref{subsubsec: theory -- optimal perturbation}) aims at finding the initial condition $\mathbf{x}_0$ that maximizes the transient amplification of energy of the response $\mathbf{x}(T) = \exp \left( \mathbfcal{A} T \right) \mathbf{x}_0$ at the target time $t=T$. It is thus an initial-value problem that can be investigated in the time domain. Rather than considering the response of the system to different initial conditions, one may instead wonder how the system reacts to external noise. For that purpose, let us now consider a forced linear dynamical system
    \begin{equation}
      \dot{\mathbf{x}} = \mathbfcal{A} \mathbf{x} + \mathbf{f}
      \label{eq: theory -- forced linear system}
    \end{equation}
    where the forcing $\mathbf{f}$ now models the system's input such as the external noise. As before, we moreover assume that all of the eigenvalues of $\mathbfcal{A}$ lie within the stable half of the complex plane. As for the optimal perturbation analysis, one may now consider a worst-case scenario, i.e.\ what is the forcing $\mathbf{f}$ that maximizes the asymptotic response of the system? Because we consider a linear dynamical system, this question can naturally be addressed in the frequency domain.

    In the most general case, the response of the system to the forcing $\mathbf{f}(t)$ is given by
    \begin{equation}
      \mathbf{x}(t) = \int_0^t \exp \left( \mathbfcal{A} (t-\tau) \right) \mathbf{f}(\tau) \ \mathrm{d}\tau
      \label{eq: theory -- convolution integral}
    \end{equation}
    which is a convolution integral. Note that, in the above expression, we assumed a zero initial condition, i.e.\ $\mathbf{x}_0 = 0$. Such a convolution integral is also known as a memory integral and highlights that the current state $\mathbf{x}(t)$ of the system depends on the entire history of the forcing $\mathbf{f}$. Because we consider linear stable systems, the influence of the forcing on the current state decays exponentially according to the least stable eigenvalue. Let us assume furthermore a harmonic external forcing
    \begin{equation}
      \mathbf{f}(t) = \Re \left( \hat{\mathbf{f}} e^{i \omega t} \right)
    \end{equation}
    where $\omega \in \mathbb{R}$ is the circular frequency of the forcing. The convolution integral can now be easily computed in the frequency domain. Given our assumptions, the asymptotic response of the system at the frequency $\omega$ is given by
    \begin{equation}
      \hat{\mathbf{x}} = \left( i \omega \mathbfcal{I} - \mathbfcal{A} \right)^{-1} \hat{\mathbf{f}}.
      \label{eq: theory -- transfer function}
    \end{equation}
    where $\hat{\mathbf{x}}$ and $\hat{\mathbf{f}}$ are the Fourier transforms of $\mathbf{x}$ and $\mathbf{f}$, respectively. The operator $\mathbfcal{R}(\omega) = \left( i \omega \mathbfcal{I} - \mathbfcal{A} \right)^{-1}$ appearing in Eq. \eqref{eq: theory -- transfer function} is known as the \emph{Resolvent operator} and is related to the exponential propagator $\mathbfcal{M}(t) = \exp \left( \mathbfcal{A} t \right)$ via Laplace transform. This operator, acting in the frequency domain, maps the input harmonic forcing $\hat{\mathbf{f}}(\omega)$ to the output harmonic response $\hat{\mathbf{x}}(\omega)$.

    Finding the forcing frequency $\omega$ that maximizes the asymptotic response $\mathbf{x}$ of the system can now be formalized as
    \begin{equation}
      \begin{aligned}
        \mathcal{R}(\omega) & = \max_{\hat{\mathbf{f}}} \displaystyle \frac{\| \left( i \omega \mathbfcal{I} - \mathbfcal{A} \right)^{-1} \hat{\mathbf{f}} \|_2^2}{\| \hat{\mathbf{f}} \|_2^2} \\
        & = \| \mathbfcal{R}(\omega) \|_2^2.
      \end{aligned}
      \label{eq: theory -- resolvent norm}
    \end{equation}
    Going from the time domain to the frequency domain, the norm of the exponential propagator is replaced with that of the resolvent in order to quantify the energy amplification between the input forcing and the output response. As before, the optimal resolvent gain at the frequency $\omega$ is given by
    \begin{equation}
      \mathcal{R}(\omega) = \sigma_1^2,
      \notag
    \end{equation}
    where $\sigma_1$ is the largest singular value of $\mathbfcal{R}(\omega)$. The associated optimal forcing $\hat{\mathbf{f}}_{\mathrm{opt}}$ and response $\hat{\mathbf{x}}_{\mathrm{opt}}$ are then given by the corresponding right and left singular vectors, respectively.

    % --> Illustration.
    \paragraph{Illustration}

    Let us illustrate resolvent analysis using the linearized complex Ginzburg-Landau equation, a typical model for instabilities in spatially-evolving flows. The equation reads
    \begin{equation}
      \displaystyle \frac{\partial u}{\partial t} = \displaystyle -\nu \frac{\partial u}{\partial x} + \gamma \frac{\partial^2 u}{\partial x^2} + \mu(x) u.
      \label{eq: theory -- ginzburg-landau equation}
    \end{equation}
    The spatial dependency of the solution result from the parameter $\mu(x)$ which is defined as
    $$\mu(x) = (\mu_0 - c_{\mu}^2) + \displaystyle \frac{\mu_2}{2} x^2.$$
    The same expression has been used in \cite{prsla:hunt:1991, amr:bagheri:2009, jfm:chen:2011}. We take $\mu_0 = 0.23$ and all other parameters are set to the same values as in \cite{amr:bagheri:2009}. The resulting model is linearly stable but is susceptible to large non-modal growth. We use the same code as \cite{jfm:chen:2011}. The problem is discretized on the interval $x \in \left[ -85, 85 \right]$ using 220 points with a pseudo-spectral approach based on Hermite polynomials.

\begin{figure}
      \centering
      \includegraphics[scale=1]{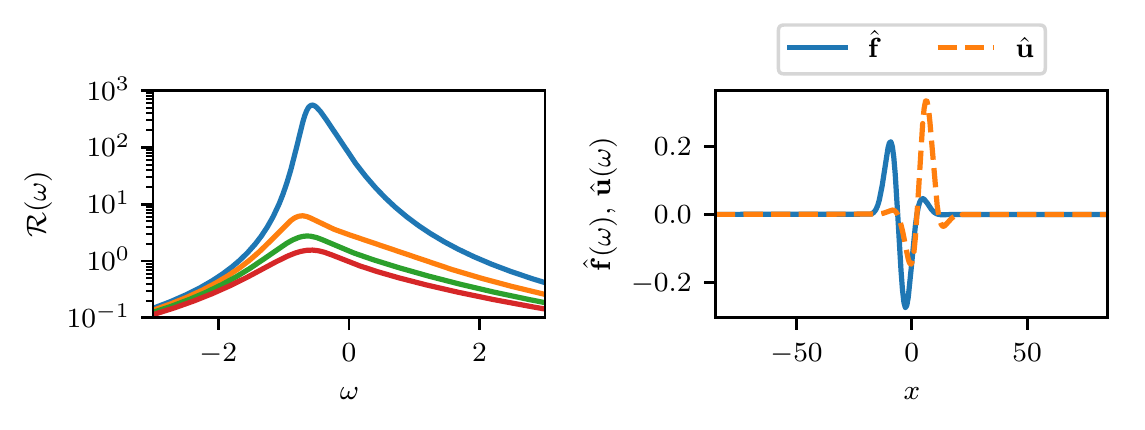}
      \caption{(left) Evolution of the first four resolvent gains $\sigma_j^2$ as a function of the forcing frequency $\omega$ for the complex Ginzburg-Landau equation \eqref{eq: theory -- ginzburg-landau equation}. (right) Optimal forcing $\hat{\mathbf{f}}(\omega)$ and associated optimal response $\hat{\mathbf{u}}(\omega)$ for the most amplified frequency. Note that only the real parts are shown.}
      \label{fig: theory -- resolvent analysis}
    \end{figure}

    Figure \ref{fig: theory -- resolvent analysis}(a) depicts the evolution of the first four resolvent gains $\sigma_j^2$ as a function of the forcing frequency $\omega$. Although the system is linearly stable for the set of parameters considered, a unit-norm harmonic forcing $\hat{\mathbf{f}}(\omega)$ can trigger a response $\hat{\mathbf{u}}(\omega)$ whose energy has been amplified by a factor almost 1000. The optimal forcing and associated response for the most amplified frequency ($\omega \simeq -0.55$) are depicted on figure \ref{fig: theory -- resolvent analysis}(b). It can be observed that their spatial support are disjoint. The optimal forcing is mostly localized in the downstream region $-20 \le x \le 0$, while the associated response is mostly localized in the upstream region $0 \le x \le 20$. This difference in the spatial support of the forcing and the response is a classical feature of highly non-normal spatially evolving flows. Such a behavior, which has been observed in a wide variety of open shear flows, has a lot of implications when it comes to flow control, see \cite{amr:bagheri:2009} for more details.

% --> Numerical methods.
\section{Numerical methods}
\label{sec: numerics}

\begin{figure}[b]
  \centering
  \includegraphics[width=.75\textwidth]{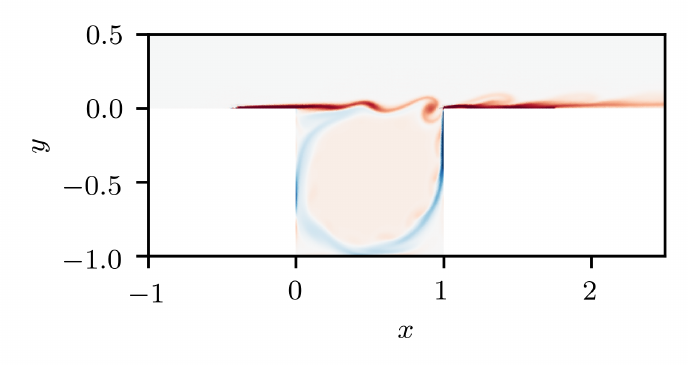}
  \caption{Instantaneous vorticity field of the shear-driven cavity flow at $Re=7500$ (based on the cavity's depth).}
  \label{fig: numerics -- shear-driven cavity flow}
\end{figure}

In this section, different techniques will be presented to solve modal and non-modal stability problems for very large-scale dynamical systems. Such very large-scale systems typically arise from the spatial discretization of partial differential equations, e.g.\ the Navier-Stokes equations in fluid dynamics. Throughout this section, the two-dimensional shear-driven cavity flow at various Reynolds numbers will serve as an example. The same configuration as \cite{jfm:sipp:2007} is considered. The dynamics of the flow are governed by
\begin{equation}
  \begin{aligned}
    \displaystyle \frac{\partial \mathbf{U}}{\partial t} + \left( \mathbf{U} \cdot \nabla \right) \mathbf{U} & = - \nabla P + \frac{1}{Re} \nabla^2 \mathbf{U} \\
    \nabla \cdot \mathbf{U} & = 0,
  \end{aligned}
  \label{eq: numerics -- Navier-Stokes equations}
\end{equation}
where $\mathbf{U}$ is the velocity field and $P$ is the pressure field. Figure \ref{fig: numerics -- shear-driven cavity flow} depicts a typical vorticity snapshot obtained from direct numerical simulation at a supercritical Reynolds number.

Given a fixed point $\mathbf{U}_b$ of the Navier-Stokes equations \eqref{eq: numerics -- Navier-Stokes equations}, the dynamics of an infinitesimal perturbation $\mathbf{u}$ evolving on top of it are governed by
\begin{equation}
  \begin{aligned}
    \displaystyle \frac{\partial \mathbf{u}}{\partial t} + \left( \mathbf{u} \cdot \nabla \right) \mathbf{U}_b  + \left( \mathbf{U}_b \cdot \nabla \right) \mathbf{u} & = - \nabla p + \frac{1}{Re} \nabla^2 \mathbf{u} \\
    \nabla \cdot \mathbf{u} & = 0.
  \end{aligned}
  \label{eq: numerics -- linearized Navier-Stokes equations}
\end{equation}
Once projected onto a divergence-free vector space, Eq. \eqref{eq: numerics -- linearized Navier-Stokes equations} can be formally written as
\begin{equation}
  \dot{\mathbf{u}} = \mathbfcal{A}\mathbf{u},
  \label{eq: numerics -- linearized Navier-Stokes equations bis}
\end{equation}
where $\mathbfcal{A}$ is the linearized Navier-Stokes operator. After being discretized in space, $\mathbfcal{A}$ is a $n \times n$ matrix. For our example, the computational domain is discretized using 158 400 grid points, resulting in a total of 475 200 degrees of freedom. From a practical point of view, explicitly assembling the resulting matrix $\mathbfcal{A}$ would have relatively large memory footprint. Using explicitly the matrix $\mathbfcal{A}$ to investigate the stability properties of this two-dimensional flow is thus hardly possible on a simple laptop at the moment despite the simplicity of the case considered. It has to be noted however that, given an initial condition $\mathbf{u}_0$, the analytical solution to Eq. \eqref{eq: numerics -- linearized Navier-Stokes equations bis} reads
\begin{equation}
  \mathbf{u}(T) = \exp \left( \mathbfcal{A}T \right) \mathbf{u}_0,
  \notag
\end{equation}
where $\mathbfcal{M} = \exp \left( \mathbfcal{A}T \right)$ is the exponential propagator introduced previously. Although assembling explicitly this matrix $\mathbfcal{M}$ is even harder than assembling $\mathbfcal{A}$, its application onto the vector $\mathbf{u}_0$ can easily be computed using a classical time-stepping code solving the linearized Navier-Stokes equations \eqref{eq: numerics -- linearized Navier-Stokes equations}. Such a \emph{time-stepper} approach has been popularized by \cite{jcp:edwards:1994, aiaa:bagheri:2009}. In the rest of this section, the different algorithms proposed for fixed point computation, linear stability and non-modal stability analyses will heavily rely on this time-stepper strategy. The key point is that they require only minor modifications of an existing time-stepping code to be put into use.

  %%%%%%%%%%%%%%%%%%%%%%%%%%%%%%%%%%%%%%%%%%%%%%%%%%%%%%%
  %%%%%                                             %%%%%
  %%%%%     KRYLOV METHODS FOR LINEAR EQUATIONS     %%%%%
  %%%%%                                             %%%%%
  %%%%%%%%%%%%%%%%%%%%%%%%%%%%%%%%%%%%%%%%%%%%%%%%%%%%%%%

  % \subsection{Krylov methods for for solving linear systems}
  % \label{subsubsec: theory -- krylov methods}

  %%%%%%%%%%%%%%%%%%%%%%%%%%%%%%%%
  %%%%%                      %%%%%
  %%%%%     FIXED POINTS     %%%%%
  %%%%%                      %%%%%
  %%%%%%%%%%%%%%%%%%%%%%%%%%%%%%%%

  \subsection{Fixed points computation}
  \label{subsec: numerics-fixed points computation}

  The starting point when investigating a nonlinear dynamical system it to determine its fixed points. As discussed in \textsection \ref{subsec: theory-fixed points}, for a continuous-time dynamical system, such points are solution to
  \begin{equation}
    \mathbfcal{F} \left( \mathbf{X} \right) = 0,
    \label{eq: numerics -- continuous-time fixed point}
  \end{equation}
  while one needs to solve
  \begin{equation}
    \mathbf{X} - \mathbfcal{G} \left( \mathbf{X} \right) = 0
    \label{eq: numerics -- discrete-time fixed point}
  \end{equation}
  for a discrete-time nonlinear dynamical system. In this section, three different fixed point solvers will be presented.

    %-----> Selective frequency damping.
    \subsubsection{Selective Frequency Damping}
    \label{subsubsec: numerics -- selective frequency damping}

    Selective frequency damping is a fixed point computation technique proposed by {\AA}kervik \emph{et al.}\ \cite{pof:akervik:2006} in 2006 and largely adapted from the original work of Pruett \emph{et al.}\ \cite{pof:pruett:2003, pof:pruett:2006} on temporal approximate deconvolution models for large-eddy simulations. It has since become one of the standard approaches for fixed point computation in fluid dynamics due to its ease of implementation. Note that various implementations of the original selective frequency damping method have been proposed over the years \cite{pof:jordi:2014, pof:jordi:2015, pof:cunha:2015}. Moreover, it has since been extented to compute steady states of the Reynolds-Averaged-Navier-Stokes (RANS) equations \cite{cf:richez:2016} as well as for the computation of unstable periodic orbits \cite{prf:leopold:2017}. In the rest of this section, only the original formulation by {\AA}kervik \emph{et al.}\ \cite{pof:akervik:2006} will be described.

    Let us consider a fixed point $\mathbf{X}^*$ of the nonlinear system
    $$\dot{\mathbf{X}} = \mathbfcal{F} \left( \mathbf{X} \right).$$
    If $\mathbf{X}^*$ is linearly unstable, then any initial condition $\mathbf{X}_0 \neq \mathbf{X}^*$ will quickly depart from $\mathbf{X}^*$. Using standard regularization techniques from control theory, the aim of selective frequency damping is thus to stabilize the linearly unstable fixed point $\mathbf{X}^*$. For that purpose, one can use proportional feedback control so that the forced system now reads
    \begin{equation}
      \dot{\mathbf{X}} = \mathbfcal{F} \left( \mathbf{X} \right) - \chi \left( \mathbf{X} - \mathbf{Y} \right),
      \label{eq: numerics -- sfd forced system}
    \end{equation}
    where $\chi$ is the control gain and $\mathbf{Y}$ the target solution. This target solution is obviously the fixed point one aims to stabilize, i.e.\ $\mathbf{Y} = \mathbf{X}^*$, which is unfortunately not known \emph{a priori}. It has to be noted however that, for a large range of situations, the instability of the fixed point $\mathbf{X}^*$ will tend to give rise to unsteady dynamics. In such cases, the target solution $\mathbf{Y}$ is thus a modification of $\mathbf{X}$ with \emph{reduced temporal fluctuations}, i.e.\ a temporally low-pass filtered solution. This filtered solution is defined as
    \begin{equation}
      \mathbf{Y}(t) = \mathcal{H}(t, \Delta) * \mathbf{X}(t-\tau)
      \label{eq: numerics -- low-pass filtered solution}
    \end{equation}
    where $\mathcal{H}$ is the convolution kernel of the applied causal low-pass filter and $\Delta$ the filter witdh. Using such definitions, the forced system \eqref{eq: numerics -- sfd forced system} can thus be rewritten as
    \begin{equation}
      \dot{\mathbf{X}} = \mathbfcal{F}\left( \mathbf{X} \right) - \chi \left( \mathbfcal{I} - \mathcal{H} \right) * \mathbf{X}.
      \label{eq: numerics -- sfd foced system bis}
    \end{equation}
    As $\mathbf{X}$ tends to the fixed point $\mathbf{X}^*$, the low-pass filtered solution $\mathbf{Y}$ tends to $\mathbf{X}$. Once a steady state has been reached, one has
    $$\mathbf{X} = \mathbf{Y} = \mathbf{X}^*,$$
    i.e.\ the fixed point of the controlled system \eqref{eq: numerics -- sfd foced system bis} is the same as that of our original system. Moreover, as the system approaches its fixed point, the amplitude of the proportional feedback control term vanishes.

\begin{figure}[b]
      \sidecaption
      \includegraphics[scale=1]{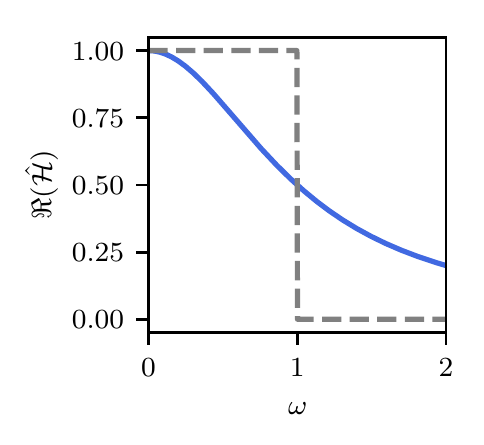}
      \caption{Evolution of $\Re \left( \hat{\mathcal{H}} \right)$ ({\color{blue} ---}), i.e.\ the real part of the Laplace transform of the exponential filter, as a function of the frequency $\omega$ for $\Delta=1$. The gray dashed line depicts the ideal spectral cutoff filter.}
      \label{fig: numerics -- lapalce transform}
    \end{figure}

    As it is formulated, computing the low-pass filtered solution \eqref{eq: numerics -- low-pass filtered solution} requires the evaluation of the following convolution integral
    \begin{equation}
      \mathbf{Y}(t) = \int_{-\infty}^t \mathcal{H}(\tau-t, \Delta) \mathbf{X}(\tau) \mathrm{d}\tau.
      \label{eq: numerics -- convolution integral}
    \end{equation}
    Note that, to be admissible, the kernel $\mathcal{H}$ must be positive and properly normalized. Moreover, in the limit of vanishing filter width, it must approach the Dirac delta function. To the best of our knowledge, all implementations of the selective frequency damping thus relies on the exponential kernel
    \begin{equation}
      \mathcal{H}(\tau - t, \Delta) = \displaystyle \frac{1}{\Delta} \exp \left( \frac{\tau - t}{\Delta} \right).
      \label{eq: numerics -- exponential kernel}
    \end{equation}
    The corresponding Laplace transform is given by
    \begin{equation}
      \hat{\mathcal{H}}(\omega, \Delta) = \displaystyle \frac{1}{1 + i \omega \Delta}.
      \label{eq: numerics -- laplace transform}
    \end{equation}
    The cutoff frequency of this filter is given by $\omega_c = \nicefrac{1}{\Delta}$. Figure \ref{fig: numerics -- lapalce transform} depicts the real part of $\hat{\mathcal{H}}$ as a function of the frequency $\omega$ for $\Delta=1$. Naturally, this cutoff frequency needs to be tuned so that the frequency associated to the instability one aims to kill is quenched by the filter.

    For real applications, evaluating the convolution integral \eqref{eq: numerics -- convolution integral} is impractical as it necessitates the storage of the complete time history of $\mathbf{X}$. Consequently, it is replaced by its differential form given by
    \begin{equation}
      \dot{\mathbf{Y}} = \displaystyle \frac{1}{\Delta} \left( \mathbf{X} - \mathbf{Y} \right)
      \label{eq: numerics -- differential filter formulation}
    \end{equation}
    which can be integrated in time using classical integration schemes, e.g.\ second-order Euler. Combining \eqref{eq: numerics -- differential filter formulation} and \eqref{eq: numerics -- sfd forced system} finally yields to the following extended system
    \begin{equation}
      \left\{
      \begin{aligned}
        \dot{\mathbf{X}} & = \mathbfcal{F}\left( \mathbf{X} \right) - \chi \left( \mathbf{X} - \mathbf{Y} \right) \\
        \dot{\mathbf{Y}} & = \displaystyle \frac{1}{\Delta} \left( \mathbf{X} - \mathbf{Y} \right).
      \end{aligned}
      \right.
      \label{eq: numerics -- selective frequency damping}
    \end{equation}
    Implementing \eqref{eq: numerics -- selective frequency damping} into an existing time-stepping code requires only minor modifications, hence making it an easy choice for fixed point computation. It must be emphasized however that, because it relies on a low-pass filtering procedure, this selective frequency damping method is unable to quench non-oscillating instabilities, e.g.\ instabilities arising due to a pitchfork bifurcation. This particular point is one of its major limitations.

    %-----> Newton-Krylov method.
    \subsubsection{Newton-Krylov methods}
    \label{subsubsec: numerics -- newton-krylov methods}

    While we relied on the continuous time representation of our system in \textsection \ref{subsubsec: numerics -- selective frequency damping}, we now turn to its discrete-time counterpart. For that purpose, consider the following nonlinear system
    \begin{equation}
      \mathbf{x}_{k+1} = \mathbfcal{G}\left( \mathbf{x}_k \right).
      \label{eq: numerics -- discrete time system}
    \end{equation}
    Our goal is thus to find a fixed point $\mathbf{x}^*$ of this system. Newton-Raphson method is a natural choice, provided the dimension of $\mathbf{x}$ is not too large. For large-scale dynamical systems, one may turn to the class of Newton-Krylov methods instead. These encompass a wide variety of different approaches, part of which have been reviewed in \cite{jcp:knoll:2004}. In the rest of this section, a variant of the recursive projection method (RPM) originally proposed by Shroff \& Keller \cite{siam:shroff:1993} will be presented.

    Iteration \eqref{eq: numerics -- discrete time system} converges if all the eigenvalues $\{ \mu_k \}_1^n$ of the Jacobian of $\mathbfcal{G}$ lie in the unit disk and the initial iterate $\mathbf{x}_0$ is sufficiently close to the actual fixed point $\mathbf{x}^*$. It will however fail even if a single eigenvalue of the Jacobian lies outside the unit disk. Note that, for our purposes, the Jacobian of $\mathbfcal{G}$ is given by the exponential propagator
    \begin{equation}
      \mathbfcal{M} = \exp \left( \mathbfcal{A} T \right).
      \label{eq: numerics -- jacobian matrix}
    \end{equation}
    The basic Newton iteration reads
    \begin{equation}
      \tilde{\mathbf{x}}_{k+1} = \tilde{\mathbf{x}}_k - \left( \mathbfcal{I} - \mathbfcal{M} \right)^{-1} \left( \tilde{\mathbf{x}}_k - \mathbfcal{G}(\tilde{\mathbf{x}}_k) \right)
    \end{equation}
    with $$\lim_{k \to +\infty} \tilde{\mathbf{x}}_k = \mathbf{x}^*,$$
    that is, as $k \to +\infty$, the Newton iterate $\tilde{\mathbf{x}}_k$ converges toward the fixed point $\mathbf{x}^*$ of the system under consideration. Note however that this Newton iteration requires the inversion of a large $n \times n$ matrix, something that may be quite impractical if $n$ is large.

    Let us now suppose that a small number $m$ of eigenvalues lies outside the disk
    $$K_{\delta} = \left\{ \vert z \vert \leq 1-\delta \right\},$$
    that is
    $$\vert \mu_1 \vert \geq \cdots \geq \vert \mu_m \vert > 1 - \delta > \vert \mu_{m+1} \vert \geq \cdots \geq \vert \mu_{n} \vert.$$
    and denote by $\mathbb{P}$ the maximal invariant subspace of $\mathbfcal{M}$ belonging to $\left\{ \mu_k \right\}_1^m$ while $\mathbb{Q}$ denotes its orthogonal complement, i.e.\ $\mathbb{P} + \mathbb{Q} = \mathbb{R}^n$. Introducing $\mathbfcal{P}$ and $\mathbfcal{Q}$ as the orthogonal projectors onto these two subspaces, we have, for each $\mathbf{x} \in \mathbb{R}^n$, the unique decomposition
    \begin{equation}
      \mathbf{x} = \mathbf{p} + \mathbf{q}, \qquad \mathbf{p} \equiv \mathbfcal{P} \mathbf{x} \in \mathbb{P}, \qquad \mathbf{q} \equiv \mathbfcal{Q} \mathbf{x} \in \mathbb{Q}.
      \label{eq: numerics -- orthogonal decomposition}
    \end{equation}
    Using these two projectors, the Lyapunov-Schmidt decomposition of Eq.\ \eqref{eq: numerics -- discrete time system} finally reads
    \begin{eqnarray}
        \label{eq: numerics -- lyapunov-schmidt unstable subspace}
        \mathbf{p}_{k+1} = \bm{f}\left( \mathbf{p}_k, \mathbf{q}_k \right) \equiv \mathbfcal{P} \mathbfcal{G}\left( \mathbf{p}_k + \mathbf{q}_k \right)\\
        \label{eq: numerics -- lyapunov-schmidt stable subspace}
        \mathbf{q}_{k+1} = \bm{g}\left( \mathbf{p}_k, \mathbf{q}_k \right) \equiv \mathbfcal{Q} \mathbfcal{G}\left( \mathbf{p}_k + \mathbf{q}_k \right).
    \end{eqnarray}
    As shown in \cite{siam:shroff:1993}, even though Eq.\ \eqref{eq: numerics -- discrete time system} may diverge, Eq.\ \eqref{eq: numerics -- lyapunov-schmidt stable subspace} is locally convergent on $\mathbb{Q}$ in the vicinity of the fixed point $\mathbf{x}^* = \mathbf{p}^* + \mathbf{q}^*$. The key idea of the \emph{recursive projection method} is thus to stabilize Eq.\ \eqref{eq: numerics -- discrete time system} by using a Newton method within the low-dimensional unstable subspace $\mathbb{P}$ while continuing to use the classical fixed-point iteration within its orthogonal complement $\mathbb{Q}$. The stabilized system then reads
    \begin{equation}
      \begin{aligned}
        & \mathbf{p}_{k+1} = \mathbf{p}_k + \left( \mathbfcal{I} - \bm{f}_p \right)^{-1} \left( \bm{f}\left(\mathbf{p}_k, \mathbf{q}_k \right) - \mathbf{p}_k \right) \\
        & \mathbf{q}_{k+1} = \bm{g}\left( \mathbf{p}_k, \mathbf{q}_k \right),
      \end{aligned}
      \label{eq: numerics -- stabilized rpm system}
    \end{equation}
    where $\bm{f}_p$ is the restriction of the Jacobian $\mathbfcal{M}$ (evaluated at the current $\mathbf{x}_k$) onto the unstable subspace $\mathbb{P}$.

    Solving directly the stabilized system \eqref{eq: numerics -- stabilized rpm system} is quite impractical as it still requires the inversion of the $n \times n$ matrix $\mathbfcal{I} - \bm{f}_p$. It must be noted however that $\bm{f}_p$ being the restriction of the Jacobian $\mathbfcal{M}$ onto the low-dimensional unstable subspace $\mathbb{P}$, it has a low-rank structure. Consequently, given an orthonormal set of vectors $\mathbfcal{U} \in \mathbb{R}^{n \times m}$ than spans $\mathbb{P}$, one can write
    \begin{equation}
      \begin{aligned}
        & \mathbf{p} = \mathbfcal{U} \mathbf{z} \\
        & \mathbf{q} = \left( \mathbfcal{I} - \mathbfcal{U}\mathbfcal{U}^T \right) \mathbf{x},
      \end{aligned}
    \end{equation}
    where $\mathbf{z} \in \mathbb{R}^m$ (with $m \ll n$) is the projection of $\mathbf{p}$ onto the span of $\mathbfcal{U}$. Different procedures have been proposed to obtain the orthonormal set of vectors $\mathbfcal{U}$. Here, we use the Arnoldi or Krylov-Schur decompositions of $\mathbfcal{M}$ described in \textsection \ref{subsubsec: numerics -- arnoldi} and \textsection \ref{subsubsec: numerics -- krylov-schur}, respectively. One major benefit of these decompositions is that they do not require explicitly the matrix $\mathbfcal{M}$ but only a function that computes the corresponding matrix-vector product.

    Starting from the original system $\mathbf{x}_{k+1} = \mathbfcal{G}(\mathbf{x}_k)$, the basic RPM update $\tilde{\mathbf{x}}_{k+1}$ can finally be expressed as
    \begin{equation}
      \tilde{\mathbf{x}}_{k+1} = \mathbf{x}_{k+1} - \mathbfcal{U}\mathbfcal{U}^T \left( \mathbf{x}_{k+1} - \tilde{\mathbf{x}}_k \right) + \mathbfcal{U} \left( \mathbf{I} - \mathbfcal{H} \right)^{-1} \mathbfcal{U}^T \left( \mathbf{x}_{k+1} - \tilde{\mathbf{x}}_k \right),
      \label{eq: numerics -- basic rpm iteration}
    \end{equation}
    where $\mathbfcal{H} = \mathbfcal{U}^T \mathbfcal{M} \mathbfcal{U}$ is the projection of the high-dimensional Jacobian matrix onto the low-dimensional unstable subspace $\mathbb{P}$. By doing so, one only needs to invert a small $m \times m$ matrix at each iteration of the Newton-RPM solver.

    Finally, looking at Eq.\ \eqref{eq: numerics -- basic rpm iteration}, RPM can be understood as a predictor-corrector. First, a new prediction $\mathbf{x}_{k+1}$ is obtained from the original system. Then, it is corrected by RPM in a two-step procedure:
    \begin{enumerate}
      \item the unstable part of the residual, $\mathbfcal{U} \mathbfcal{U}^T(\mathbf{x}_{k+1} - \tilde{\mathbf{x}}_k)$, is subtracted from the predicted iterate $\mathbf{x}_{k+1}$,

      \item it is then replaced by its Newton correction, $\mathbfcal{U} \left( \mathbf{I} - \mathbfcal{H} \right)^{-1} \mathbfcal{U} \left( \mathbf{x}_{k+1} - \tilde{\mathbf{x}}_k \right)$, hence resulting in the new RPM iterate $\tilde{\mathbf{x}}_{k+1}$.
    \end{enumerate}
    Although the present fixed-point computation strategy requires substantially more modifications of an existing time-stepper solver than the selective frequency damping procedure described in \textsection \ref{subsubsec: numerics -- selective frequency damping}, it nonetheless has a number of key benefits. First, while selective frequency damping cannot compute the linearly unstable fixed point of a system if the associated instability is non-oscillating (i.e.\ associated to a real eigenvalue), the recursive projection method can. More importantly, the recursive projection method improves its approximation of the unstable subspace of the Jacobian matrix $\mathbfcal{M}$ at each iteration. Consequently, as the procedure converges to the fixed point $\mathbf{x}^*$, one obtains as a by-product really good approximations of the leading eigenvalues and associated eigenvectors of the matrix $\mathbfcal{M}$. Finally, the recursive projection method can relatively easily be extended to compute linearly unstable periodic orbits or to perform branch continuation. For more details about RPM and illustrations, interested readers are referred to \cite{siam:shroff:1993, appmaths:janovsky:2003, aiaa:campobasso:2004, jcp:renac:2011} and references therein.

    %-----> BoostConv.
    \subsubsection{BoostConv}
    \label{subsubsec: numerics -- BoostConv}

    The Newton-Krylov method presented in \textsection \ref{subsubsec: numerics -- newton-krylov methods} is a valid alternative to selective frequency damping (see \textsection \ref{subsubsec: numerics -- selective frequency damping}), particularly if a steady bifurcation occurs so that a non-oscillatory instability needs to be quenched as to recover the unstable stationary solution. Nevertheless, implementing RPM is not straightforward as it requires the Jacobian matrix (or a good approximation) of the discrete-time system considered. Recently, \cite{citro2017efficient} have introduced \emph{BoostConv}, a new fixed point computation technique somehow related to the recursive projection method.

    Let us rewrite Eq.\ \eqref{eq: numerics -- discrete time system} as
    \begin{equation}
      \mathbf{x}_{k+1} = \mathbf{x}_k + \mathbf{r}_k
      \label{eq: numerics -- boostconv discrete system}
    \end{equation}
    where $\mathbf{r}_k$ is the residual vector produced at each nonlinear iteration. It can be shown that the residual at time $k$ and time $k+1$ are related by
    \begin{equation}
      \mathbf{r}_{k+1} \simeq \mathbf{r}_k - \mathbfcal{C} \mathbf{r}_k
      \label{eq: numerics: boostconv -- residual equation}
    \end{equation}
    where $\mathbfcal{C}$ is an unknown linear operator approximately governing the dynamics of the residual vector. The key idea of \emph{BoostConv} is to define a corrected residual vector $\boldsymbol{\xi}_{k}$ such that the above equation becomes
    \begin{equation}
      \mathbf{r}_{k+1} \simeq \mathbf{r}_k - \mathbfcal{C} \boldsymbol{\xi}_k.
      \label{eq: numerics -- boostconv discrete residual equation}
    \end{equation}
    Clearly, the residual $\mathbf{r}_{k+1}$ is annihilated if one has
    \begin{equation}
      \mathbfcal{C} \boldsymbol{\xi}_k = \mathbf{r}_k.
      \label{eq: numerics -- boostconv weird equation}
    \end{equation}
    Let us now consider the two Krylov sequence of residuals
    \begin{equation*}
      \mathbfcal{X} = \text{span}\{ \mathbf{r}_k\}_1^m \quad \text{ and } \quad \mathbfcal{Y} = \text{span}\{ \mathbf{r}_k-\mathbf{r}_{k+1}\}_1^m
    \end{equation*}
    so that
    $$\mathbfcal{Y} \simeq \mathbfcal{C} \mathbfcal{X}.$$
    Assuming that the corrected residual $\boldsymbol{\xi}_k$ is a linear combination of the previous residuals stored in $\mathbfcal{X}$, the least-square solution to Eq.\ \eqref{eq: numerics -- boostconv weird equation} is given by
    \begin{equation}
      \boldsymbol{\xi}_k = \mathbfcal{X} \mathbfcal{Y}^{\dagger} \mathbf{r}_k,
    \end{equation}
    where $\mathbfcal{Y}^{\dagger}$ is the Moore-Penrose pseudoinverse. Introducing this least-square solution into Eq.\ \eqref{eq: numerics -- boostconv discrete residual equation} yields the modified residual at time $k+1$
    \begin{equation}
      \tilde{\mathbf{r}}_{k+1} = (\mathbfcal{I}-\mathbfcal{Y}\mathbfcal{Y}^{\dagger})\mathbf{r}_k.
      \label{eq: numerics -- residual k+1 w/o correction}
    \end{equation}
    Looking at the above equation, our least-square trick thus allows us to annihilate the residual within the subspace defined by the column span of $\mathbfcal{Y}$ while leaving it untouched in the orthogonal complement. Piecing everything together, the stabilized BoostConv iterate can finally be written as
    \begin{equation}
      \tilde{\mathbf{x}}_{k+1} = \mathbf{x}_{k+1} - \mathbfcal{Y} \mathbfcal{Y}^{\dagger}\left( \mathbf{x}_{k+1} - \tilde{\mathbf{x}}_k \right) + \mathbfcal{X} \mathbfcal{Y}^{\dagger} \left( \mathbf{x}_{k+1} - \tilde{\mathbf{x}}_k \right).
    \end{equation}
    Just like the recursive projection method, BoostConv can be understood as a predictor-corrector iterative scheme. First, a new prediction $\mathbf{x}_{k+1}$ is obtained from the original system. Then, it is corrected by BoostConv in a two-step procedure:
    \begin{enumerate}
      \item the unstable part of the residual $\mathbfcal{YY}^{\dagger}\left( \mathbf{x}_{k+1} - \tilde{\mathbf{x}}_k \right)$ is first subtracted from the prediction $\mathbf{x}_{k+1}$
      \item it is then replaced by its least-square correction $\mathbfcal{XY}^{\dagger} \left( \mathbf{x}_{k+1} - \tilde{\mathbf{x}}_k \right)$ which plays the same role as the low-dimensional Newton correction step in RPM.
    \end{enumerate}
    Although RPM and BoostConv appear closely related, the latter does not require the Jacobian matrix $\mathbfcal{M}$ nor an estimate of it. It can moreover be implemented as a black box around an existing solver since it only requires the residual $\mathbf{r}_k$ as input and returns the corrected one $\boldsymbol{\xi}_k$ as output. Also, BoostConv can be easily adapted to stabilize periodic orbits \cite{citro2017efficient}.

    %-----> Comparison.
    \subsubsection{Comparison of the different approaches}
    \label{subsubsec: numerics -- fixed points comparison}

    Computing the linearly unstable fixed point of the Navier-Stokes equations for the two-dimensional shear-driven cavity flow at $Re=4150$ provides a typical benchmark to illustrate the performances of the three fixed points solvers presented, namely \emph{selective frequency damping} (see \textsection \ref{subsubsec: numerics -- selective frequency damping}), the \emph{recursive projection method} (see \textsection \ref{subsubsec: numerics -- newton-krylov methods}) and \emph{BoostConv} (see \textsection \ref{subsubsec: numerics -- BoostConv}). The different methods have been setup as follows:
    \begin{itemize}
      \item \emph{Selective frequency damping}: the cutoff frequency of the low-pass filter has been set to $\omega_c = 3.5$ while the gain is set to $\chi = 0.15$. These parameters, chosen based on trial and errors, provide the best performances for SFD that we have observed.

      \item \emph{Recursive projection method}: the dimension of the Krylov subspace providing the orthonormal basis for the leading invariant subspace of the Jacobian matrix has been set to $k_{\mathrm{dim}} = 10$. The outer RPM iteration $\mathbf{x}_{k+1} = \mathbfcal{G}(\mathbf{x}_k)$ has been setup so that it corresponds to 100 time-steps of the Navier-Stokes solver.

      \item \emph{BoostConv}: it has been parametrized as RPM.
    \end{itemize}
    Each method iterates until the norm of the Navier-Stokes solver's residual is below $\epsilon = 10^{-10}$. Figure \ref{fig: numerics -- comparison of sfd, boostconv, rpm} depicts the vorticity field of the linearly unstable solution to the Navier-Stokes equations computed by the recursive projection method. Although not shown, the other two methods converge toward the same unstable equilibrium solution. The evolution of the residual as a function of the number of iterations performed by the nonlinear Navier-Stokes solver is reported in figure \ref{fig: numerics -- comparison of sfd, boostconv, rpm bis}. It appears quite clearly that the recursive projection method largely outperforms the selective frequency damping and BoostConv. On the other hand, BoostConv appears only marginally more efficient than the selective frequency damping procedure. This comparison is however biased as it does not include the computational cost of constructing the orthonormal projection basis needed in the RPM solver. Given how similar BoostConv and RPM are, this plot nonetheless highlights the importance of correctly approximating the leading unstable subspace of the Jacobian matrix. Although it has been partially addressed in the original paper \cite{siam:shroff:1993} and in \cite{jcp:renac:2011}, this particular point currently focuses our efforts.

    \begin{figure}[b]
      \sidecaption
      \centering
      \includegraphics[scale=1]{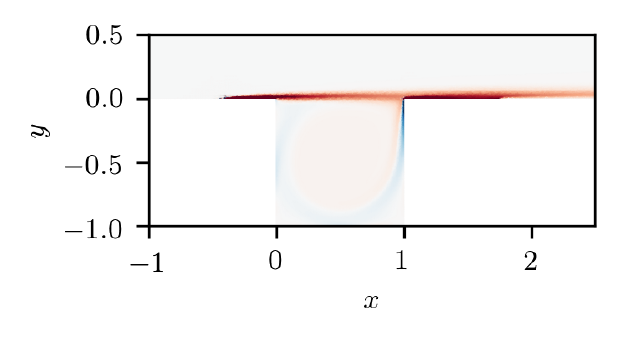}
      \caption{Vorticity field of the linearly unstable solution to the Navier-Stokes equations obtained by the recursive projection method. Red denotes negative vorticity (i.e.\ clockwise rotation) while blue denotes positive vorticity (i.e.\ counter clockwise rotation).}
      \label{fig: numerics -- comparison of sfd, boostconv, rpm}
    \end{figure}

    \begin{figure}[b]
      \centering
      \includegraphics[scale=1]{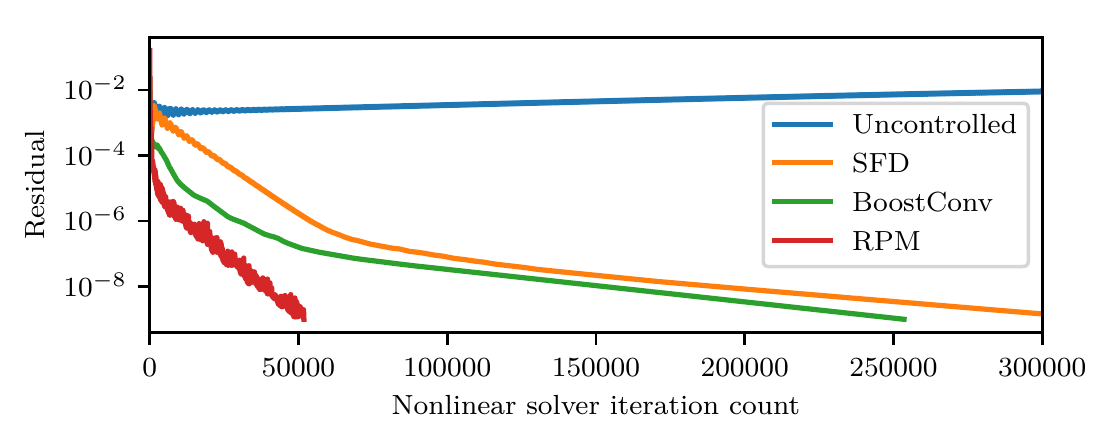}
      \caption{Comparison of the residual evolution obtained by selective frequency damping (see \textsection \ref{subsubsec: numerics -- selective frequency damping}), the recursive projection method (see \textsection \ref{subsubsec: numerics --  newton-krylov methods}) and BoostConv (see \textsection \ref{subsubsec: numerics -- BoostConv}). The evolution of the uncontrolled residual is also reported for the sake of completeness. The benchmark considered is that of the two-dimensional shear-driven cavity flow at $Re = 4150$.}
      \label{fig: numerics -- comparison of sfd, boostconv, rpm bis}
    \end{figure}

  %%%%%%%%%%%%%%%%%%%%%%%%%%%%%%%%%%%%%%%%%%%%
  %%%%%                                  %%%%%
  %%%%%     MODAL STABILITY ANALYSIS     %%%%%
  %%%%%                                  %%%%%
  %%%%%%%%%%%%%%%%%%%%%%%%%%%%%%%%%%%%%%%%%%%%

  \subsection{Linear stability and eigenvalue computation}

  The aim of linear stability analysis is to determine whether a perturbation $\textbf{x}$, governed by
  \begin{equation}
    \dot{ \mathbf{x} } = \mathbfcal{A} \mathbf{x},
    \notag
  \end{equation}
  will grow or decay exponentially rapidly as $t \to \infty$. This asymptotic behavior is entirely governed by the eigenspectrum of the Jacobian matrix $\mathbfcal{A}$: if at least one of its eigenvalues has a positive (resp. negative) real part, the linear system considered is unstable (resp. stable), see \textsection \ref{subsec: theory -- linear stability} for more details.

  It must be emphasized that, within a time-stepper framework, one does not seek directly for the eigenpairs of the Jacobian matrix $\mathbfcal{A}$ of the continuous-time problem. Instead, the problem considered is recast in the discrete-time framework as
  \begin{equation}
    \mathbf{x}_{k+1} = \mathbfcal{M} \mathbf{x}_k,
    \label{eq: numerics -- discrete-time linear system}
  \end{equation}
  where $\mathbfcal{M} = \exp \left( \mathbfcal{A} T \right)$ is the exponential propagator already introduced in \textsection \ref{subsec: theory -- linear stability}, \textsection \ref{subsubsec: theory -- optimal perturbation}, and \textsection \ref{subsubsec: numerics -- newton-krylov methods}, and where $T$ is the sampling period. The system is then linearly unstable if at least one eigenvalue $\mu$ of $\mathbfcal{M}$ lies outside the unit disk, i.e. $\vert \mu \vert > 1$.

  As discussed previously, although one cannot explicitly assemble the exponential propagator $\mathbfcal{M}$, its action onto a given vector $\mathbf{x}_k$ simply amounts to march in time the linearized system from $t = kT$ to $t = (k+1)T$. This ability to evaluate relatively easily the matrix-vector product given by \eqref{eq: numerics -- discrete-time linear system} allows us to use iterative solvers in order to compute the eigenpairs of $\mathbfcal{M}$. The rest of this section is thus devoted to the presentation of two iterative eigenvalue solvers, namely the Arnoldi decomposition and the Krylov-Schur decomposition.

    %-----> Arnoldi decomposition.
    \subsubsection{Arnoldi decomposition}
    \label{subsubsec: numerics -- arnoldi}

    Let us denote the following sequence of vectors
    \begin{equation}
      \mathbfcal{K}_m\left( \mathbfcal{M}, \mathbf{v}_0 \right) = \left\{ \mathbf{v}_0, \mathbfcal{M}\mathbf{v}_0, \cdots, \mathbfcal{M}^{m-1} \mathbf{v}_0 \right\}.
      \label{eq: numerics -- krylov sequence}
    \end{equation}
    Eq. \eqref{eq: numerics -- krylov sequence} is known as a \emph{Krylov sequence}. It eventually converges toward the eigenvector associated to the largest eigenvalue (in modulus) of $\mathbfcal{M}$ as $m \to \infty$. Generating this sequence to approximate the leading eigenpair of $\mathbfcal{M}$ is known as the \emph{power iteration method}. Note that this simple method retains only the last vector of this sequence while discarding the information contained in the first $m-1$ vectors.

\begin{figure}[b]
      \centering
      \includegraphics[width=.9\textwidth]{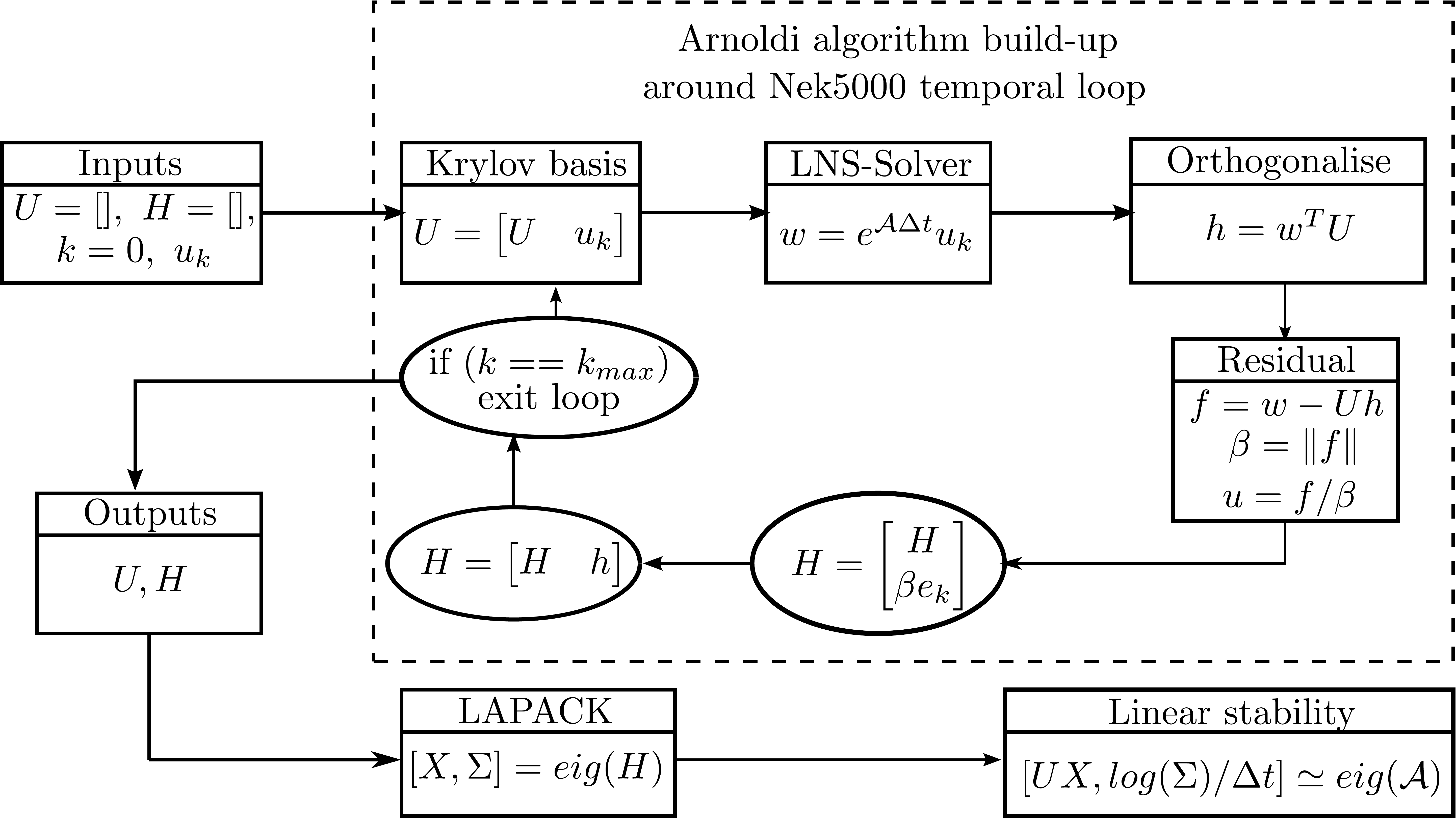}
      \caption{Block-diagram representation of the basic $m$-step Arnoldi factorization. Note that, within a time-stepper framework, every matrix-vector product $\mathbfcal{M} \mathbf{v}_i$ is evaluated by marching in time the linearized system considered.}
      \label{fig: numerics -- arnoldi decomposition}
    \end{figure}

    Contrary to the power iteration method, Arnoldi decomposition uses all of the information contained in the Krylov sequence \eqref{eq: numerics -- krylov sequence} as to compute better estimates of the leading eigenvalues of $\mathbfcal{M}$. Readers can easily be convinced that the Krylov sequence \eqref{eq: numerics -- krylov sequence} obeys
    \begin{equation}
      \mathbfcal{M} \mathbfcal{K}_m \simeq \mathbfcal{K}_m \mathbfcal{C},
      \notag
    \end{equation}
    where $\mathbfcal{C}$ is a $m \times m$ companion matrix representing the low-dimensional projection of $\mathbfcal{M}$ onto the span of the Krylov sequence \eqref{eq: numerics -- krylov sequence}. As such, the eigenpairs of $\mathbfcal{C}$ approximate the leading eigenpairs of $\mathbfcal{M}$. It must be emphasized however that, as $m$ increases, the last vectors in the Krylov sequence become almost parallel. Consequently, the companion matrix $\mathbfcal{C}$ becomes increasingly ill-conditioned. In order to overcome the loss of information from the power iteration method and the increasingly ill-conditioned companion matrix decomposition, the Arnoldi method combines them with a Gram-Schmidt orthogonalization process. The basic Arnoldi iteration then reads
    \begin{equation}
      \mathbfcal{MV}_m = \mathbfcal{V}_m \mathbfcal{H}_m + \mathbf{r}_{m} \mathbf{e}^T_m,
      \label{eq: numerics -- m-step Arnoldi}
    \end{equation}
    where $\mathbfcal{V}_m$ is an orthonormal set of vectors, $\mathbfcal{H}_m$ is a $m \times m$ upper Hessenberg matrix and $\vert \mathbf{r}_{m} \mathbf{e}^T_m \vert$ is the residual indicating how far $\mathbfcal{V}_m$ is from from an invariant subspace of $\mathbfcal{M}$. Because of its relatively small dimension, the eigenpairs $\left( \mu_H, \mathbf{y} \right)$ of the Hessenberg matrix, also known as Ritz pairs, can be computed using direct eigensolvers. The Ritz pairs of $\mathbfcal{H}_m$ are related to the eigenpairs of $\mathbfcal{M}$ as follows
    \begin{equation}
      \begin{aligned}
        \mu_{\mathbfcal{M}} & \simeq \mu_{\mathbfcal{H}} \\
        \hat{\mathbf{u}} & \simeq \mathbfcal{V}_m \mathbf{y}.
      \end{aligned}
    \end{equation}
    A detailed presentation of the basic $m$-step Arnoldi factorization is given in algorithm \eqref{algo: Arnoldi} while figure \ref{fig: numerics -- arnoldi decomposition} depicts its block-diagram representation to ease the understanding. As can be seen, Arnoldi decomposition is relatively simple to implement within an existing time-stepper code. One has to bear in mind however that, in order to capture (within a time-stepper framework) an eigenpair of the Jacobian matrix $\mathbfcal{A}$ characterized by a circular frequency $\omega$, one has to obey the Nyquist criterion and needs at least four snapshots to appropriately discretize the associated period.

    \begin{algorithm}
      \begin{algorithmic}
        \REQUIRE $\mathbfcal{M} \in \mathbb{R}^{n \times n}$, starting vector $\mathbf{v} \in \mathbb{R}^n$.\\
        $\mathbf{v}_1 = \mathbf{v} / \| \mathbf{v} \|$;\\
        $\mathbf{w} = \mathbfcal{M} \mathbf{v}_1$; \\
        $\alpha_1 = \mathbf{v}_1^T \mathbf{w}$;\\
        $\mathbf{f}_1 \leftarrow \mathbf{w} - \alpha_1 \mathbf{v}_1$;\\
        $\mathbfcal{V}_1 \leftarrow \left( \mathbf{v}_1 \right)$; \\
        $\mathbfcal{H}_1 \leftarrow (\alpha_1)$;\\
        \FOR{$j=1,2,\cdots,m-1$}
        \STATE{$\beta_j=\|\mathbf{f}_j\|$; \\
        $\mathbf{v}_{j+1} \leftarrow \mathbf{f}_j/\beta_j$; \\
        $\mathbfcal{V}_{j+1} \leftarrow \left( \mathbfcal{V}_j, \mathbf{v}_{j+1} \right)$;\\

        $\hat{\mathbfcal{H}}_j \leftarrow \begin{pmatrix}
                                            \mathbfcal{H}_j \\
                                            \beta_j \mathbf{e}_j^T
                                          \end{pmatrix}
        $ \\
        $\mathbf{w} \leftarrow \mathbfcal{M} \mathbf{v}_{j+1}$;\\
        $\mathbf{h} \leftarrow \mathbfcal{V}_{j+1}^T \mathbf{w}$; \\
        $\mathbf{f}_{j+1} \leftarrow \mathbf{w} - \mathbfcal{V}_{j+1} \mathbf{h}$;\\
        $\mathbfcal{H}_{j+1} \leftarrow (\hat{\mathbfcal{H}}_j,\mathbf{h})$;}
        \ENDFOR
      \end{algorithmic}
      \caption{The $m$-step \emph{Arnoldi} factorisation.}
      \label{algo: Arnoldi}
    \end{algorithm}

    \begin{figure}[b]
      \centering
      \includegraphics[width=.75\textwidth]{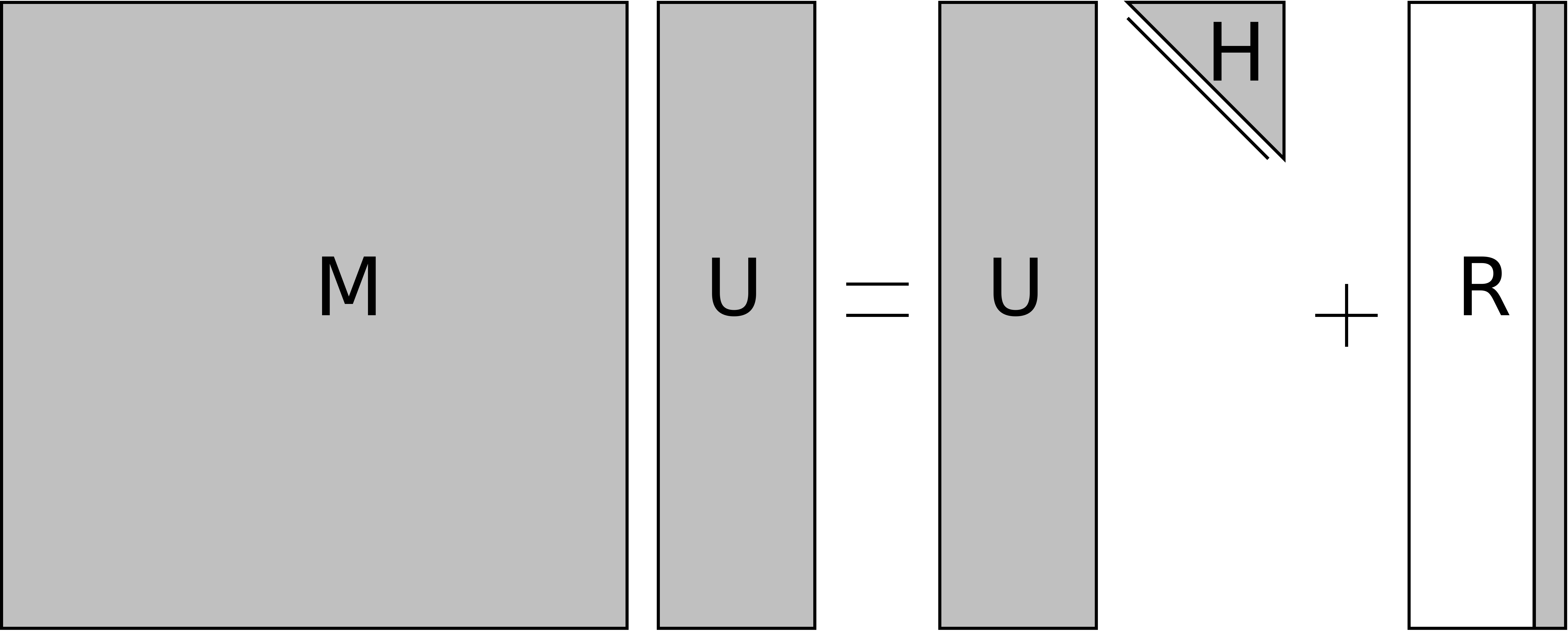}
      \caption{\emph{Arnoldi decomposition} -- Given a matrix $\mathbfcal{M} \in \mathbb{R}^{n \times n}$, construct an orthonormal set of vectors $\mathbfcal{V} \in \mathbb{R}^{n \times k}$ such that $\mathbfcal{H} \in \mathbb{R}^{k \times k}$ is an upper Hessenberg matrix and only the last column of the residual matrix $\mathbfcal{R} \in \mathbb{R}^{n \times k}$ is nonzero. Figure has been adapted from \cite{book:antoulas:2005}.}
      \label{fig: numerics -- arnoldi decomposition structure}
    \end{figure}

    %-----> Krylov-Schur decomposition.
    \subsubsection{Krylov-Schur decomposition}
    \label{subsubsec: numerics -- krylov-schur}

    Let us consider the $m$-step Arnoldi factorization
    \begin{equation}
      \mathbfcal{MV}_m = \mathbfcal{V}_m \mathbfcal{H}_m + \beta \mathbf{v}_{m+1} \mathbf{e}^T_m
      \label{eq: m-step Arnoldi}
    \end{equation}
    introduced in \textsection \ref{subsubsec: numerics -- arnoldi}. As discussed previously, the Ritz pair $\left( \mu_H, \mathbfcal{V}_m \mathbf{y} \right)$ of $\mathbfcal{H}_m$ provides a good approximation for the eigenpair $\left( \mu, \hat{\mathbf u} \right)$ of the matrix $\mathbfcal{M}$. One limitation of the Arnoldi decomposition is however that the dimension $m$ of the Krylov subspace necessary to converge the leading Ritz pairs is not known \emph{a priori}. It might hence be relatively large, thus potentially causing some numerical and/or practical problems (e.g.\ storage of Krylov basis $\mathbfcal{V}_m$, forward instability of the Gram-Schmidt process involved in the Arnoldi decomposition, etc). Two different approaches have been proposed to overcome these limitations: the \emph{Implicitly Restarted Arnoldi Method} introduced by Sorensen \cite{Sorensen_SIAM_1992} in 1992 and the \emph{Krylov-Schur decomposition} introduced by Stewart \cite{Stewart_SIAM_2001} in 2001. In the present work, the latter approach has been preferred because of its simplicity of implementation and its robustness.

    The Krylov-Schur method is based on the generalization of the m-step Arnoldi factorization~\eqref{eq: m-step Arnoldi} to a \emph{Krylov decomposition} of order $m$
    \begin{equation}
      \mathbfcal{MV}_m = \mathbfcal{V}_m \mathbfcal{B}_m + \mathbf{v}_{m+1} \mathbf{b}_{m+1}^T
      \label{eq: Krylov decomposition}
    \end{equation}
    for which the matrix $\mathbfcal{B}_m$ and the vector $\mathbf{b}_{m+1}$ have no restriction. The Arnoldi decomposition then appears as a special case of Krylov decomposition where $\mathbfcal{B}_m$ is restricted to be in upper Hessenberg form and $\mathbf{b}_{m+1} = \mathbf{e}_m$. Another special case is the \emph{Krylov-Schur} decomposition for which the matrix $\mathbfcal{B}_m$ is in real Schur form (i.e.\ quasi-triangular form with its eigenvalues in the $1 \times 1$ or $2 \times 2$ diagonal blocks). It has been shown by Stewart \cite{Stewart_SIAM_2001} that Krylov and Arnoldi decompositions are equivalent (i.e.\ they have the same Ritz approximations). Moreover, by means of orthogonal similarity transformations, any Krylov decomposition can be transformed into an equivalent Krylov-Schur decomposition.

    The core of the Krylov-Schur method is based on a two-steps procedure: (\emph{i}) an expansion step performed using a $m$-step Arnoldi factorization, and (\emph{ii}) a contraction step to a Krylov-Schur decomposition of order $p$ retaining only the most useful spectral information from the initial $m$-step Arnoldi decomposition. Given an initial unit-norm vector $\mathbf{v}_1$, a subroutine to compute the matrix-vector product $\mathbfcal{M} \mathbf{v}_i$, and the desired dimension $m$ of the Krylov subspace, the Krylov-Schur method can be summarized as follows:
    \begin{enumerate}
      \item Construct an initial Krylov decomposition of order $m$ using for instance the $m$-step Arnoldi factorization~\eqref{eq: m-step Arnoldi}.

      \item Check for the convergence of the Ritz eigenpairs. If a sufficient number has converged, then stop. Otherwise, proceed to step 3.

      \item Compute the real Schur decomposition $\mathbfcal{B}_m = \mathbfcal{Q} \mathbfcal{S}_m \mathbfcal{Q}^T$ such that the matrix $\mathbfcal{S}_m$ is in real Schur form and $\mathbfcal{Q}$ is the associated matrix of Schur vectors. It is assumed furthermore that the Ritz values on the diagonal blocks of $\mathbfcal{S}_m$ have been sorted such that the $p$ ''wanted'' Ritz values are in the upper-left corner of $\mathbfcal{S}_m$, while the $m-p$ ''unwanted'' ones are in the lower-right corner. At this point, we have the following re-ordered Krylov-Schur decomposition
      \begin{equation}
        \mathbfcal{M} \tilde{\mathbfcal{V}}_m =
        \tilde{\mathbfcal{V}}_m
        \begin{bmatrix}
         \mathbfcal{S}_{11} & \mathbfcal{S}_{12} \\
         {\mathbf 0}     & \mathbfcal{S}_{22}
       \end{bmatrix}
       + {\mathbf v}_{m+1}\begin{bmatrix}
                           {\mathbf b}_{1}^T & {\mathbf b}_{2}^T
                           \end{bmatrix}
        \label{eq: Krylov-Schur decomposition}
      \end{equation}
      with $\tilde{\mathbfcal{V}}_m = \mathbfcal{V}_m  \mathbfcal{Q}$ being the re-ordered Krylov basis, $\mathbfcal{S}_{11}$ the subset of the Schur matrix containing the $p$ ''wanted'' Ritz values, $\mathbfcal{S}_{22}$ the subset containing the $m-p$ ''unwanted'' ones, and $\begin{bmatrix} {\mathbf b}_1^T & {\mathbf b}_2^T \end{bmatrix} = {\mathbf b}^T \mathbfcal{Q}$.

      \item Truncate the Krylov-Schur decomposition~\eqref{eq: Krylov-Schur decomposition} of order $m$ to a Krylov decomposition of order $p$,
        \begin{equation}
          \mathbfcal{M}\tilde{\mathbfcal{V}}_p = \tilde{\mathbfcal{V}}_p \mathbfcal{S}_{11} + \tilde{\mathbf v}_{p+1}{\mathbf b}_1^T
        \end{equation}
      with $\tilde{\mathbfcal{V}}_p$ equal to the first $p$ columns of $\tilde{\mathbfcal{V}}_m$ and $\tilde{\mathbf v}_{p+1} = {\mathbf v}_{m+1}$.

      \item Extend again to a Krylov decomposition of order $m$ using a variation of the procedure used in the first step: the procedure is re-initialized with the starting vector ${\mathbf v}_{p+1}$ but all the vectors in $\tilde{\mathbfcal{V}}_p$ are taken into account in the orthogonalization step.

      \item Check the convergence of the Ritz values. If not enough Ritz values have converged, restart from step 3.

    \end{enumerate}
    This algorithm has two critical steps. The first one is the choice of the ''wanted'' Ritz values in the re-ordering of the Schur decomposition in step 2. Since we are only interested in the leading eigenvalues of the linearized Navier-Stokes operator, all the Ritz pairs being classified as ''wanted'' must satisfy $\left| \mu_w \right| \ge 1 - \delta$ (with $\delta = 0.05 - 0.1$ usually). Regarding the criterion assessing the convergence of a given Ritz pair, starting from the Krylov decomposition~\eqref{eq: m-step Arnoldi}, one can write
    \begin{equation}
      \| \mathbfcal{M} \mathbfcal{V}_m \mathbf{y} - \mathbfcal{V}_m \mathbfcal{B}_m \mathbf{y} \| = \| \mathbfcal{M} \mathbfcal{V}_m \mathbf{y} - \mu_{\mathbf B} \mathbfcal{V}_m \mathbf{y} \| = \left| \beta \mathbf{e}_m^T \mathbf{y} \right|
      \label{eq: Krylov convergence}
    \end{equation}
    with $(\mu_{\mathbf B},\mathbf{y})$ a given eigenpair of the matrix $\mathbfcal{B}_m$. If the right hand side $\left| \beta {\mathbf e}_{m}^T{\mathbf y} \right|$ is smaller than a given tolerance, then the Ritz pair $(\mu_{\mathbf B}, \mathbfcal{V}_m {\mathbf y})$ provides a good approximation to the eigenpair $(\mu, \hat{\mathbf u})$ of the original matrix $\mathbfcal{M}$. A Ritz value is generally considered as being converged if the associated residual $\left| \beta {\mathbf e}_{m}^T{\mathbf y} \right| \le 10^{-6}$.

    %-----> Comparisons.
    \subsubsection{Comparison of the two approaches}
    \label{subsubsec: numerics -- comparison arnoldi krylov-schur}

    Following the comparison of the fixed points solvers in \textsection \ref{subsubsec: numerics -- fixed points comparison}, let us now compare the efficiency of the time-stepper Krylov-Schur decomposition over the Arnoldi one when computing the leading eigenvalues and eigenmodes of the linearized Navier-Stokes operator for the shear-driven cavity flow at $Re=4150$. For that purpose, the eigenspectrum obtained using the Arnoldi decomposition with a Krylov subspace of dimension $k_{\mathrm{dim}}=256$ will serve as our reference point. For the sake of comparison, three Krylov subspaces of various dimensions, namely $k_{\mathrm{dim}}=$192, 128 and 64 have been considered for the Krylov-Schur decomposition. In all cases, twelve eigenvalues were required to have converged with a residual $\epsilon \le 10^{-6}$ before the computation could stop. Finally, the sampling period has been set to $T = 0.2$ non-dimensional time units so that the exponential propagator (whose action is approximated by time-marching the linearized Navier-Stokes equation) is given by $\mathbfcal{M} = \exp \left( 0.2 \mathbfcal{A} \right)$, with $\mathbfcal{A}$ being the linearized Navier-Stokes operator.

\begin{figure}
      \centering
      \includegraphics[scale=1]{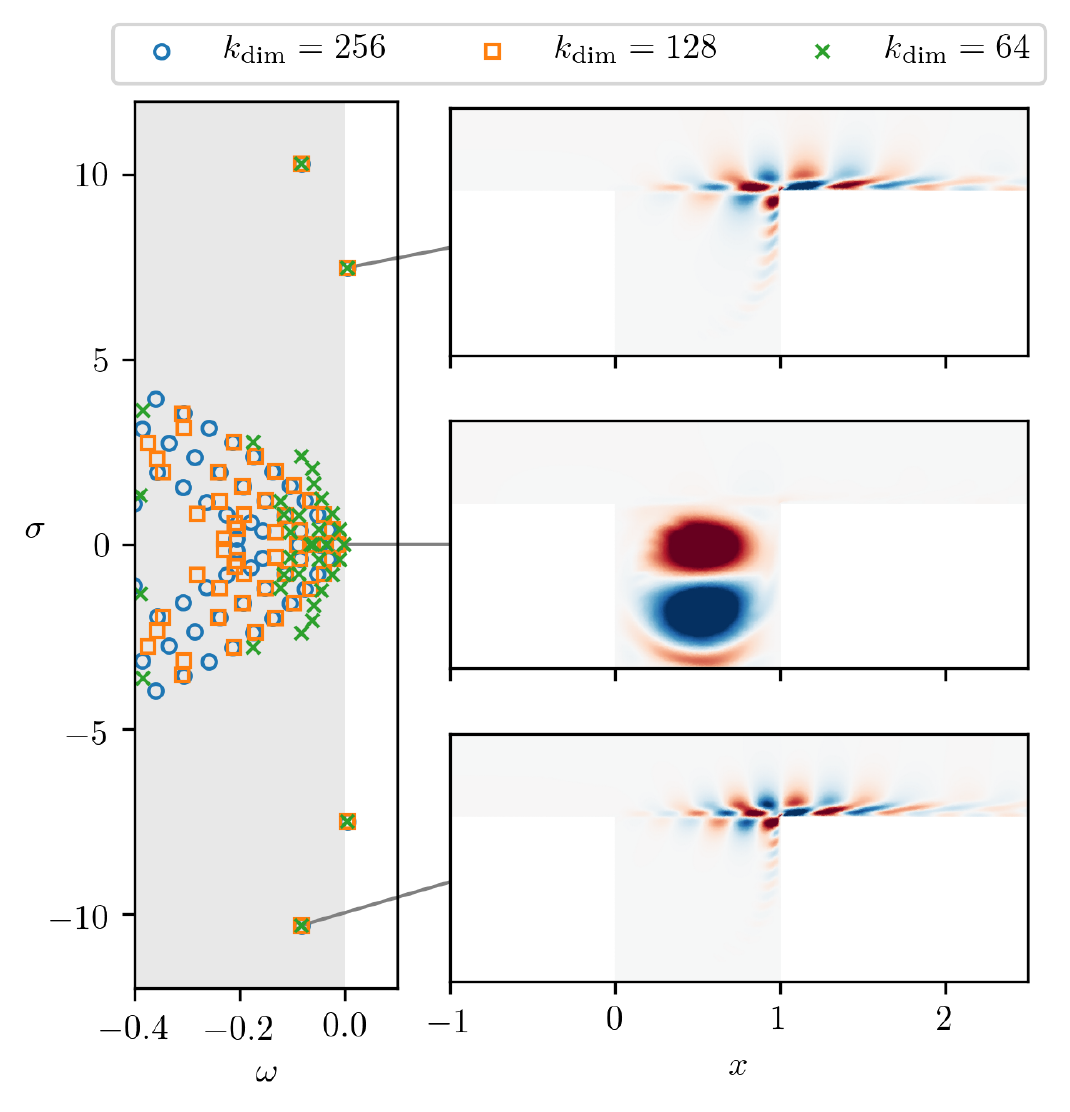}
      \caption{Eigenspectrum and leading eigenmodes (streamwise velocity field) for the shear-driven cavity flow at $Re=4150$. Blue circles ({\color{blue} $\boldsymbol{\circ}$}) depict the eigenvalues obtained using the Arnoldi decomposition (see \textsection \ref{subsubsec: numerics -- arnoldi}) with a Krylov subspace of dimension $k_{\mathrm{dim}} = 256$ while orange squares ({\color{orange} $\square$}) and green crosses ({\color{green} $\times$}) depict the eigenvalues obtained using the Krylov-Schur decomposition (see \textsection \ref{subsubsec: numerics -- krylov-schur}) with a Krylov subspace of dimension $k_{\mathrm{dim}}=128$ and $k_{\mathrm{dim}} = 64$, respectively. In both cases, the computation stopped once the twelve eigenvalues have been converged down to $\epsilon = 10^{-6}$.}
      \label{fig: numerics -- linear stability eigenpairs}
    \end{figure}

    Left panel of figure \ref{fig: numerics -- linear stability eigenpairs} depicts the eigenspectra obtained using the Arnoldi decomposition with a Krylov subspace dimension $k_{\mathrm{dim}}=256$ and Krylov-Schur decomposition with $k_{\mathrm{dim}}=128$ and $k_{\mathrm{dim}}=64$, while its right panel shows the real part of the streamwise velocity component of some of the leading eigenmodes for the sake of completeness. These plots highlight the existence of two families of modes: (\textit{i}) high-frequency shear layer modes (also known as \emph{R\"ossiter} modes), and (\textit{ii}) low-frequency inner-cavity modes similar to the ones existing in lid-driven cavities. A detailed description of the physical mechanisms underlying these instabilities is beyond the scope of the present work. Interested readers are referred to \cite{BLLFD2011} regarding the shear layer instability modes and \cite{FALP2007,FPLFB2009} for the inner-cavity instabilities.

    Table \ref{tab: numerics -- linear stability comparison} reports the growth rate $\sigma$ and circular frequency $\omega$ of the leading eigenvalue for all of the cases considered. Even with a Krylov subspace four times smaller than the reference one, it can be seen that the leading eigenvalue's growth rate computed by the Krylov-Schur decomposition differs by less than 0.02\% compared to our reference solution while the circular frequency is left unchanged at least until the fifth digit. The major difference between $k_{\text{dim}} = 256$ and $k_{\text{dim}}=64$ is the accuracy of the eigenvalues belonging to the branches of inner-cavity modes, see figure \ref{fig: numerics -- linear stability eigenpairs}. These modes however turn out to be of limited interest in the dynamics of the flow.

    Finally, the last row of table \ref{tab: numerics -- linear stability comparison} reports the total number of calls to the linearized Navier-Stokes solver necessary to converge the required twelve eigenvalues. Compared to our reference case (i.e.\ $k_{\text{dim}}=256$), the total number of matrix-vector multiplications is inversely proportional to the reduction factor of the dimension of the Krylov subspace considered. Despite this increase of the number of matrix-vector multiplications, Krylov-Schur decomposition with a moderate subspace dimension (e.g.\ $k_{\text{dim}}=64$ or 128) has nonetheless two major benefits compared to the classical Arnoldi decomposition:
    \begin{itemize}
      \item \emph{Reduced memory footprint}: having a smaller Krylov subspace implies that fewer Krylov vectors need to be stored. This may be of crucial importance if one consider very large-scale eigenvalue problems such as the ones appearing in fluid dynamics (see \cite{jfm:ilak:2012, jfm:loiseau:2014, pof:citro:2015, jfm:bucci:2018} for illustrations) which need to be solved on high-performance computers.

      \item \emph{Partially reduced computation complexity}: although table \ref{tab: numerics -- linear stability comparison} underlines that a larger number of calls to the linearized time-stepper code is required as we decrease the size of the Krylov subspace, one must not overlook that Arnoldi decomposition necessitates modified Gram-Schmidt orthogonalization of the Krylov sequence to iteratively construct the upper Hessenberg matrix. For an $n \times k$ matrix (where $n$ is the number of degrees of freedom and $k$ the dimension of the Krylov subspace), the computational complexity of this step scales as $nk^2$. As a consequence, decreasing the size of the Krylov subspace by a factor 4 reduces the computational cost of the modified Gram-Schmidt orthogonalization by a factor 16. Such a reduction becomes particularly attractive if ones needs a very large Krylov subspace to converge the leading eigenvalues when using the classical Arnoldi decomposition.
    \end{itemize}
    Finally, although Krylov-Schur decomposition does have some benefits compared to the classical Arnoldi iteration, it must not be forgotten that the overall computational time is dictated by the linearized time-stepper solver used to evaluate the application of the exponential propagator $\mathbfcal{M}$ onto a given vector. Consequently, efficient and scalable temporal integrators are key enablers for very large-scale eigenvalue analysis arising from the discretization of partial differential equations. Discussion on efficient and scalable temporal and/or spatial discretization is however beyond the scope of this contribution.

    \begin{table}
      \caption{Growth rate $\sigma$ and circular frequency $\omega$ of the leading eigenvalue computed for different dimensions of the Krylov subspace. Note that only the largest Krylov subspace (i.e.\ $k_{\mathrm{dim}}=256$) uses the basic Arnoldi decomposition. All other computations have been performed with Krylov-Schur. The total number of calls to the linearized Navier-Stokes solver (i.e.\ the number of Jacobian matrix-vector multiplications) is also reported for each case.}
      \label{tab: numerics -- linear stability comparison}
      \begin{tabular}{p{2cm}p{2.5cm}p{2.5cm}p{2.5cm}p{2.5cm}}
      \hline\noalign{\smallskip}
       & $k_{\mathrm{dim}} = 256$ & $k_{\mathrm{dim}} = 192$ & $k_{\mathrm{dim}} = 128$ & $k_{\mathrm{dim}} = 64$  \\
      \noalign{\smallskip}\svhline\noalign{\smallskip}
      $\sigma$  & $4.56757 \cdot 10^{-3}$ & $4.56757 \cdot 10^{-3}$ & $4.56757 \cdot 10^{-3}$ & $4.56673 \cdot 10^{-3}$\\
      $\omega$ & $\pm 7.4938$ & $\pm 7.4938$ & $\pm 7.4938$ & $\pm 7.4938$\\
      Matrix-vector multiplications & 256 & 384 & 512 & 832\\
      \noalign{\smallskip}\hline\noalign{\smallskip}
      \end{tabular}
    \end{table}
  %%%%%%%%%%%%%%%%%%%%%%%%%%%%%%%%%%%%%%%%%%%%%%%%
  %%%%%                                      %%%%%
  %%%%%     NON-MODAL STABILITY ANALYSIS     %%%%%
  %%%%%                                      %%%%%
  %%%%%%%%%%%%%%%%%%%%%%%%%%%%%%%%%%%%%%%%%%%%%%%%

  \subsection{Non-modal stability and singular value decomposition}

  Given the linear time-invariant dynamical system
  \begin{equation}
    \dot{\mathbf{x}} = \mathbfcal{A} \mathbf{x} + \mathbf{f},
    \notag
  \end{equation}
  it has been shown in \textsection \ref{subsec: theory -- linear stability} that, for $\mathbf{f} = \mathbf{0}$ (i.e.\ no external forcing), the asymptotic fate of a random initial condition $\mathbf{x}_0$ is dictated by the eigenpairs of the Jacobian matrix $\mathbfcal{A}$. On the other hand, as shown in \textsection \ref{subsec: theory -- non-modal stability}, its short-term dynamics are governed by the singular triplets of the exponential propagator $\mathbfcal{M} = \exp \left( \mathbfcal{A} T \right)$, where $T$ is the time horizon considered. Conversely, the asymptotic response of the (stable) system to an external forcing $\mathbf{f}$ is governed by the singular triplets of the so-called resolvent operator $\mathbfcal{R} = \left( i\omega \mathbfcal{I} - \mathbfcal{A} \right)^{-1}$, where $\omega$ is the forcing frequency. The rest of this section is devoted to the presentation of two different time-stepper algorithms for the computation of the leading singular values and singular vectors of the exponential propagator $\mathbfcal{M}$ or the resolvent operator $\mathbfcal{R}$.

  %-----> Optimization point of view.
  \subsubsection{An optimization approach}

\begin{figure}
    \centering
    \includegraphics[width=\textwidth]{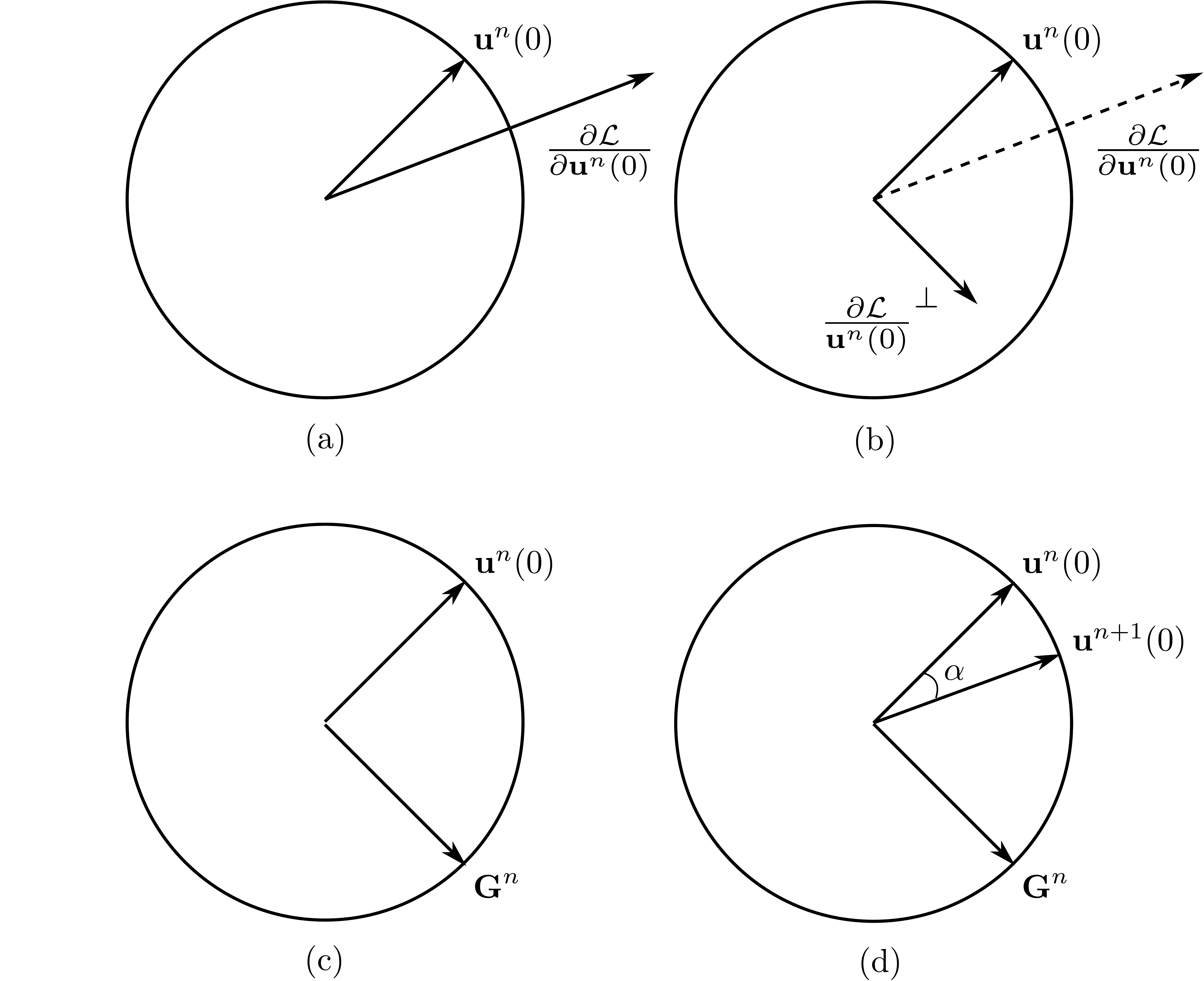}
    \caption{Schematic representation of the \emph{rotation update} method. (a) Compute the gradient $\nicefrac{\partial \mathcal{L}}{\partial \mathbf{x}}$ of the augmented Lagrange functional. (b) Orthogonalise the gradient with respect to $\mathbf{x}^n(0)$. (c) Compute $\mathcal{G}^n$, i.e.\ the orthogonalized gradient normalized such that its energy is $E_0$. (d) Update $\mathbf{x}^{n+1}(0)$ using a linear combination $\mathbf{x}^n(0)$ and of the orthonormalized gradient $\mathbf{G}^n$.}
    \label{fig: numerics -- rotation update gradient}
  \end{figure}

  It has been shown in \textsection \ref{subsec: theory -- non-modal stability} that optimal perturbation analysis could be formulated as an optimization problem given by
  \begin{equation}
      \begin{aligned}
        \maximize \limits_{\mathbf{x}_0} & \mathcal{J} \left( \mathbf{x}_0 \right) = \| \mathbf{x}(T) \|_2^2\\
        \subjecto & \dot{\mathbf{x}} - \mathbfcal{A}\mathbf{x} = 0 \\
        ~ & \| \mathbf{x}_0 \|_2^2 - 1 = 0,
      \end{aligned}
      \label{eq: numerics -- optimal perturbation constrained maximization}
  \end{equation}
  where $\mathcal{J}(\mathbf{x}_{0})$ is known as the \emph{objective function}. Similarly, the optimal forcing problem can be formulated as
  \begin{equation}
      \begin{aligned}
        \maximize \limits_{\hat{\mathbf{f}}} & \mathcal{J} \left(    \hat{\mathbf{f}} \right) = \| \hat{\mathbf{x}}(\omega) \|_2^2\\
        \subjecto & \left( i \omega \mathbfcal{I} - \mathbfcal{A} \right) \hat{\mathbf{x}} = \hat{\mathbf{f}} \\
        ~ & \| \hat{\mathbf{f}} \|_2^2 - 1 = 0,
      \end{aligned}
      \label{eq: numerics -- optimal forcing constrained maximization}
  \end{equation}
  Though these optimization problems are non-convex, solutions to both of them can be obtained by means of standard gradient-based optimization algorithms. One of the most famous such algorithms is the \emph{conjugate gradient} method originally introduced by \cite{book:hestenes:1952}, see \cite{book:saad:2003} and \cite{book:golub:2012} for more recent presentations. In this work we will however introduce the reader to the \emph{rotation update} technique, a modification of the classical steepest ascent method based on geometric considerations. Figure \ref{fig: numerics -- rotation update gradient} provides a schematic description of this algorithm. This approach has been used by \cite{jfm:foures:2013, jfm:foures:2014} and \cite{fdr:farano:2016} in the context of \emph{p-norms} optimization in fluid dynamics.

  Both Eq.\ \eqref{eq: numerics -- optimal perturbation constrained maximization} and Eq.\ \eqref{eq: numerics -- optimal forcing constrained maximization} are constrained maximization problems. As shown in \textsection \ref{subsec: theory -- non-modal stability}, introducing Lagrange multipliers allows us to transform these constrained problems into equivalent unconstrained ones. For the optimal perturbation analysis, the unconstrained maximization problem thus reads
  \begin{equation}
    \maximize \limits_{\mathbf{x}, \mathbf{v}, \mu} \mathcal{L} \left( \mathbf{x}, \mathbf{v}, \mu \right),
    \label{eq: numerics -- unconstrained maximization}
  \end{equation}
  where
  \begin{equation}
    \mathcal{L} \left( \mathbf{x}, \mathbf{v}, \mu \right) = \mathcal{J}\left( \mathbf{x}_0 \right) + \int_{0}^T \mathbf{v}^T \left( \dot{\mathbf{x}} - \mathbfcal{A}\mathbf{x} \right) \mathrm{d}t + \mu \left( \| \mathbf{x}_0 \|_2^2 - 1 \right)
    \label{eq: numerics -- augmented Lagrangian}
  \end{equation}
  is known as the \emph{augmented Lagrangian} function. The additional optimization variables $\mathbf{v}$ and $\mu$ appearing in the definition of the augmented Lagrangian $\mathcal{L}$ are the \emph{Lagrange multipliers}. The gradient of the augmented Lagrange functional $\mathcal{L}$ with respect to the initial condition $\mathbf{x}_0$ reads
  \begin{equation}
    \displaystyle \frac{\partial \mathcal{L}}{\partial \mathbf{x}_0} = 2\mu\mathbf{x}_0 - \mathbf{v}_0.
  \end{equation}
  This expression explicitly depends on the Lagrange multiplier $\mu$ whose value is unfortunately unknown. One can however write down a mathematical expression of this gradient orthogonalized with respect to the input $\mathbf{x}_0$
  \begin{equation}
      \frac{\partial \mathcal{L}}{\partial {\bf x}}^{\perp} = \frac{\partial \mathcal{L}}{\partial {\bf x}} - \frac{\langle \frac{\partial \mathcal{L}}{\partial {\bf x}},{\bf x} \rangle}{\langle {\bf x},{\bf x} \rangle}{\bf x}
  \end{equation}
  where $\langle \cdot,\cdot \rangle$ stands for the inner product. Note moreover that we dropped the subscript 0 for the sake of simplicity. It can now be expressed as
  \begin{equation}
      \frac{\partial \mathcal{L}}{\partial {\bf x}}^{\perp} = ({\bf v} - 2\mu {\bf x}) - \frac{\langle ({\bf v} - 2\mu {\bf x}) , {\bf x} \rangle}{\langle {\bf x},{\bf x} \rangle}{\bf x}
  \end{equation}
  After simplifications, the orthogonalized gradient finally reads
  \begin{equation}
    \frac{\partial \mathcal{L}}{\partial {\bf x}}^{\perp} = \mathbf{v} - \frac{\langle \mathbf{v},{\bf x} \rangle}{\langle {\bf x},{\bf x} \rangle}{\bf x}
    \label{eq: orthogonalised gradient}
  \end{equation}
  This expression now solely depends on the direct variable ${\bf x}$ and the adjoint one ${\bf v}$, while the dependence on the unknown Lagrange multiplier $\mu$ has been completely removed from the optimization problem. Normalizing this new gradient such that
  \begin{equation}
    {\bf G}^n = \sqrt{\frac{\| \mathbf{x}_0 \|_2^2}{\langle \frac{\partial \mathcal{L}}{\partial {\bf x}}^{\perp},\frac{\partial \mathcal{L}}{\partial {\bf x}}^{\perp} \rangle}} \frac{\partial \mathcal{L}}{\partial {\bf x}}^{\perp}
  \end{equation}
  now allows us to look for the update ${\bf x}^{n+1}$ as a simple linear combination of ${\bf x}^n$ and ${\bf G}^n$ given by
  \begin{equation}
    {\bf x}^{n+1} = \cos(\alpha) {\bf x}^n + \sin(\alpha) {\bf G}^n
  \end{equation}
  Since ${\bf x}^n$ and ${\bf G}^n$ form an orthonormal set of vectors, the update ${\bf x}^{n+1}$ fulfills, directly by construction, the constraint on the amplitude of the initial perturbation. No quadratic equation in $\mu$, as in the case of steepest ascent method, need to be solved anymore at each iteration of the optimization loop. To ensure the convergence of the method to the maxima of the augmented functional $\mathcal{L}$, a check needs however to be put on the value of the angle $\alpha$ used for the update of the solution. Every calculations presented in this work uses $\alpha=0.5$ as the initial value. If the value of the cost function $\mathcal{J}$ computed at the $(n+1)$\textsuperscript{th} iteration is smaller than the value of $\mathcal{J}$ at the previous one, then the update ${\bf x}^{n+1}$ is re-updated with a different value of $\alpha$, typically $\alpha = \alpha/2$ until the new value of $\mathcal{J}$ is larger than the previous one.

  %-----> Formulation as an SVD problem.
  \subsubsection{Singular value decomposition}

  Formulating the optimal perturbation analysis as a maximization problem yields a non-convex optimization problem. As a consequence, one cannot rule out the possibility that gradient-based algorithms get stuck on a local maximum. Hopefully, it has been shown in \textsection \ref{subsec: theory -- non-modal stability} that the same problem could be formulated as a singular value decomposition of either the exponential propagator $\mathbfcal{M} = \exp \left( \mathbfcal{A} T \right)$ or the resolvent one $\mathbfcal{R} = \left( i \omega \mathbfcal{I} - \mathbfcal{A} \right)^{-1}$, depending on the problem considered.

  Let us consider once more the optimal perturbation problem for the sake of simplicity, although the derivation to be given naturally extend to the case to the resolvent. Given the singular value decomposition of the exponential propagator
  \begin{equation}
    \mathbfcal{M} \triangleq \exp \left( \mathbfcal{A} T \right) = \mathbfcal{U} \boldsymbol{\Upsigma} \mathbfcal{V}^H,
    \notag
  \end{equation}
  the optimal perturbation at $t=0$ is given by the right singular vector $\mathbf{v}_1$, i.e.\ the first column of $\mathbfcal{V}$, while the associated response at time $t=T$ is given by the rescaled left singular vector $\sigma_1 \mathbf{u}_1$, where $\sigma_1$ is the associated singular value characterizing the amplification of the perturbation. Computing directly the singular values and singular vectors of $\mathbfcal{M}$ is a challenging task for very large scale problems. Hopefully, introducing the adjoint exponential propagator $\mathbfcal{M}^{\dagger} = \exp \left( \mathbfcal{A}^{\dagger} T \right)$, readers can easily be convinced that our problem can be recast as the following equivalent eigenvalue problems
  \begin{equation}
    \mathbfcal{M}^{\dagger} \mathbfcal{M} \mathbf{v} = \sigma^2 \mathbf{v} \quad \text{and} \quad \mathbfcal{M} \mathbfcal{M}^{\dagger} \mathbf{u} = \sigma^2 \mathbf{u}.
    \label{eq: numerics -- singular value decomposition as eigenvalue problem}
  \end{equation}
  From a practical point of view, evaluating the action of the matrix $\mathbfcal{M}^{\dagger} \mathbfcal{M}$ onto a vector $\mathbf{x}$ can be computed in a two-step procedure:
  \begin{enumerate}
    \item Integrate forward in time the original system, $\mathbf{x}(T) = \exp \left( \mathbfcal{A} T \right) \mathbf{x}$.
    \item Integrate backward in time the adjoint problem using the output of the previous step as the initial condition, i.e.\ evaluating $\exp \left( \mathbfcal{A}^{\dagger} T \right) \mathbf{x}(T)$.
  \end{enumerate}
  Provided an adjoint time-stepper code is available, one can thus readily solve the optimal perturbation problem using the eigenvalue solvers described in \textsection \ref{subsubsec: numerics -- arnoldi} or \textsection \ref{subsubsec: numerics -- krylov-schur}. Moreover, given that $\mathbfcal{M}^{\dagger} \mathbfcal{M}$ is a symmetric positive-definite matrix, the upper Hessenberg matrix in the Arnoldi iteration can be replaced by a tri-diagonal matrix, hence resulting in the so-called \emph{Lanczos} iteration \cite{L1950}. Note finally that, when applied to the resolvent operator $\mathbfcal{R}$, the matrix-vector product $\mathbfcal{R}(\omega) \hat{\mathbf{f}}$ can be evaluated as follows:
  \begin{enumerate}
    \item Integrate forward in time the system $\dot{\mathbf{x}} = \mathbfcal{A} \mathbf{x} + \mathbf{f}(\omega)$.
    \item Perform a discrete Fourier transform of the asymptotic response to obtain $\hat{\mathbf{u}}(\omega)$.
  \end{enumerate}
  The same procedure applies for the action of $\mathbfcal{R}^{\dagger}(\omega)$ where one now needs to integrate backward in time the adjoint system using $\hat{\mathbf{u}}(\omega)$ as the external forcing. For more details about the computation of the optimal forcing using a time-stepper approach, readers are referred to \cite{jfm:monokrousos:2010}.

  %-----> Illustration
  \subsubsection{Illustration}

\begin{figure}
    \centering
    \includegraphics[scale=1]{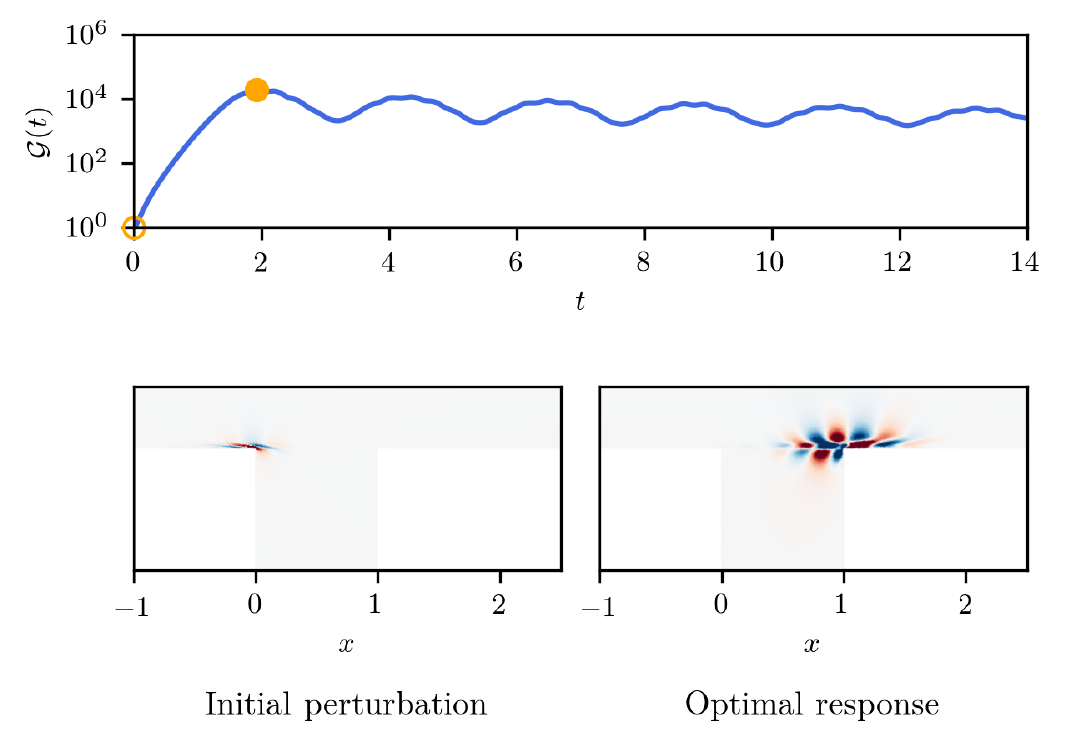}
    \caption{Time-evolution of the optimal perturbation's kinetic energy for the shear-driven cavity flow at $Re=4100$, i.e.\ below the critical Reynolds number for the onset of linear instability. The lower panels depict the streamwise velocity of the initial optimal perturbation (left) and associated optimal response (right) at $T=2$.}
    \label{fig: numerics -- illustration optimal perturbation analysis}
  \end{figure}

  As for modal stability (see \textsection \ref{subsubsec: numerics -- comparison arnoldi krylov-schur}), let us illustrate non-modal stability on the shear-driven cavity flow. For that purpose, the Reynolds number is set to $Re=4100$, i.e.\ slightly below the critical Reynolds number for the onset of linear instability. Only the optimal perturbation analysis (time-domain) will be presented for the sake of simplicity. For more details about the resolvent analysis (frequency domain), readers are referred to \cite{jfm:monokrousos:2010, phd:bucci:2017}. Figure \ref{fig: numerics -- illustration optimal perturbation analysis} depicts the evolution in time of the optimal perturbation's kinetic energy. It can be seen that, although linear stability analysis predicts that the flow is stable, perturbations can be amplified by 4 to 5 orders of magnitude solely through non-modal effects. Once the perturbation has reached its maximum transient amplification at $t=2$, its fate is eventually dictated by the least stable eigenvalue of the Jacobian matrix. Note that in the present case, the Reynolds number considered being slightly below that for the onset of instability, the eventual decay rate of the perturbation is relatively small. The perturbation nonetheless eventually disappears at $t \to +\infty$. Figure \ref{fig: numerics -- illustration optimal perturbation analysis} also clearly illustrates the different spatial support of the optimal initial perturbation (left panel) and the associated optimal response at $t=2$ (right panel). These different spatial supports result from the strong convective effects, which are related mathematically to the degree of non-normality of the Jacobian matrix. Similar observation holds true regarding the optimal forcing and optimal response when performing a resolvent analysis. Analysis of these transient (non-normal) effects may be of crucial importance when studying subcritical transition or for control purposes.

  Let us finally conclude this section by presenting the pros and cons of the SVD and optimization approaches. As discussed earlier, formulating the optimal perturbation and optimal forcing analyses in an optimization framework results in a non-convex optimization problem typically solved using gradient-based algorithms. Consequently, one cannot rule out the possibility that the solution returned by the optimization procedure actually corresponds to a local maxima of the problem at hand. On the other hand, formulating these two problems as singular value decompositions of the appropriate operator ensure that the solution obtained is indeed the optimal one by virtue of the Eckart-Young theorem. Moreover, singular value decomposition allows us to compute in one go not only the optimal perturbation but also the sub-optimal ones, something hardly possible within a classic optimization framework. Nonetheless, the optimization formulation offers much more flexibility than simply computing the optimal perturbation in the $\ell_2$ sense. Indeed, one can choose the objective function $\mathcal{J}(\mathbf{x})$ and the associated constraints according to the specific problem he/she aims to solve, see for instance \cite{jfm:foures:2013, jfm:foures:2014, fdr:farano:2016} for optimization based on the $\ell_1$ norm of the perturabtion.

% --> Conclusion and perspectives.
\section{Conclusions and perspectives}
\label{sec: conclusion}

With the ever increasing computational power available (roughly 20 to 25\% increase annually) and the development of high-performances computing (HPC), investigating the properties of realistic very large-scale nonlinear dynamical systems has become reachable. In the field of fluid dynamics, computation of fixed points of two-dimensional flows and investigation of the spectral properties of the corresponding linearized Navier-Stokes operator are now routinely performed on workstations or even laptops. The traditional way to do so is to use a so-called \emph{matrix-forming} approach where the Jacobian matrix of the system is explicitly assembled, whether one aims at computing a fixed point of the equations using Newton-like methods or at computing its leading eigenvalues and eigenmodes characterizing the linear stability properties of the fixed point considered. It must be noted however that the memory capabilities of computers increase at a slower rate than their computational capabilities. As a consequence, while simulations of very large-scale systems can now be performed (see \cite{rasthofer2017large} where the three-dimensional Navier-Stokes equations have been discretized using $10^{13}$ cells), using a matrix-forming approach to compute fixed points and study the stability properties of such systems becomes rapidly intractable. This gap between CPU and memory performances sprung the development of a new class of algorithms known as \emph{matrix-free}.

In this chapter, the reader has been introduced to a number of such matrix-free algorithms for the computation of fixed points and eigenpairs of the linearized operator. Most of these algorithms rely on the observation that existing simulation codes do not solve explicitly the continuous-time problem
$$\dot{\mathbf{x}} = \mathbfcal{F}(\mathbf{x})$$
but rather its discrete-time counterpart
$$\mathbf{x}_{k+1} = \mathbfcal{G}(\mathbf{x}_k).$$
Moreover, these time-stepper simulation codes do not form explicitly the matrices but only require to be able to compute their applications onto a given set of vectors. Given this observation, only minor modifications of existing time-stepper codes are required as to transform them into black- functions evaluating matrix-vector products, hence enabling practitioners to wrap them into powerful matrix-free iterative fixed points and/or eigenvalues solvers. Once again in the field of fluid dynamics, such an approach proved successful and allowed \cite{jfm:ilak:2012, jfm:loiseau:2014, pof:citro:2015, jfm:bucci:2018} to investigate the stability properties of the fixed points of nonlinear dynamical systems characterized by almost 50 millions degrees of freedom.

Time-stepper matrix-free approaches nonetheless suffer from a number of limitations and drawbacks. First and foremost, these approaches rely onto an existing simulation code to emulate the matrix-vector products required. As a consequence, the overall performances of the fixed points and eigenvalue solvers presented herein are essentially dictated by the efficiency of the time-stepper code considered. Second, a number of analyses (non-modal stability, receptivity, sensitivity, ...) may require the definition of an adjoint. While such adjoint operator simply reduces to the transconjugate operation in matrix-forming approaches, a dedicated adjoint time-stepper solver needs to be developed within the matrix-free framework. Although this may be quite challenging if the system is defined by a complicated set of nonlinear partial differential equations, one must note that recent developments in automatic differentiation might prove helpful (see the software TAPENADE \cite{TapenadeRef13} for instance). Finally, because the eigenvalue solvers described herein rely on Krylov techniques, one must bear in mind that only the leading subset of eigenpairs of the Jacobian matrix can be accurately computed.

Despite these limitations, time-stepper matrix-free approaches offer a practical and efficient computing framework for the investigation of very large-scale nonlinear dynamical systems. Provided one has access to an efficient time-stepper solver, their relative ease of implementation make the approaches described in the present chapter a standard choice of tools whenever matrix-forming approaches are intractable. Finally, these approaches can easily be combined with finite-differences approximation of the Jacobian matrix-vector product whenever a linearized time-stepper solver is not available, hence proving extremely versatile and applicable a very broad class of systems.

\bibliography{bibliography}
\bibliographystyle{plain}

\end{document}